\documentclass[prodmode,acmtecs]{acmsmall} 

\usepackage[ruled]{algorithm2e}

\usepackage{mathtools}
\usepackage{subcaption}
\usepackage[show]{chato-notes}
\SetAlFnt{\small}
\SetAlCapFnt{\small}
\SetAlCapNameFnt{\small}
\SetAlCapHSkip{0pt}
\IncMargin{-\parindent}

\newcommand\blfootnote[1]{%
  \begingroup
  \renewcommand\thefootnote{}\footnote{#1}%
  \addtocounter{footnote}{-1}%
  \endgroup
}

\acmYear{2019}

\setcopyright{rightsretained}

\doi{0000001.0000001}

\issn{1234-56789}

\begin{document}

\markboth{Feyisetan and Simperl}{Beyond monetary incentives}

\title{Beyond monetary incentives: experiments in paid microtask contests modelled as continuous-time markov chains}
\author{OLUWASEYI FEYISETAN
\affil{University of Southampton*}
ELENA SIMPERL
\affil{University of Southampton}}

\begin{abstract}
In this paper, we aim to gain a better understanding into how paid microtask crowdsourcing could leverage its appeal and scaling power by using contests
to boost crowd performance and engagement. We introduce our microtask-based annotation platform \textit{Wordsmith}, which features incentives such as points, leaderboards and badges on top of financial remuneration. 
Our analysis focuses on a particular type of incentive, contests, as a means to apply crowdsourcing in near-real-time scenarios, in which requesters need labels quickly. We model crowdsourcing contests as a continuous-time Markov chain with the objective to maximise the output of the crowd workers, while varying a parameter which determines whether a worker is eligible for a reward based on their present rank on the leaderboard. We conduct empirical experiments in which crowd workers recruited from CrowdFlower carry out annotation microtasks on Wordsmith - in our case, to identify named entities in a stream of Twitter posts. In the experimental conditions, we test different reward spreads and record the total number of annotations received. We compare the results against a control condition in which the same annotation task was completed on CrowdFlower without a time or contest constraint.
The experiments show that rewarding only the best contributors in a live contest could be a viable model to deliver results faster, though quality might suffer for particular types of annotation tasks. Increasing the reward spread leads to more work being completed, especially by the top contestants.
Overall, the experiments shed light on possible design improvements of paid microtasks platforms to boost task performance and speed, and make the overall experience more fair and interesting for crowd workers.
\end{abstract}

%
%


%
%

\terms{Crowdsourcing}


\acmformat{Oluwaseyi Feyisetan and Elena Simperl. X. Beyond monetary incentives: experiments in paid microtask contests.}


%

\maketitle

\section{Introduction}
\label{sec:introduction}
\blfootnote{* \emph{prior to} \textsc{Amazon Research, Seattle, USA} (\emph{work completed at the University of Southampton, UK})}More than a decade after the term was introduced by Jeff Howe in 2006 \cite{howe2006rise}, crowdsourcing seems to be considered by many as a, if not `the', universal means to solve virtually any kind of problem, online and offline, that requires sustained human involvement. We see it used to motivate employees to engage with less rewarding work routines, attract the best possible ideas to boost innovation, enhance artificial intelligence algorithms, and support ambitious social and entrepreneurial initiatives. Achieving a goal by collecting contributions from many individuals has a long tradition, far beyond the most recent developments in the digital world. In fact, some of the most successful exemplars of the `wisdom of the crowds' in modern times, for instance Wikipedia, predate Howe's article. And yet, with the rise of social networks, smart mobile devices and online platforms, the phenomenon has found a new dimension in terms of scale and achievements - it routinely manages to mobilise very large groups of people in a relatively short period of time, helping organisations from tech and government to the military and marketing improve the ways they operate, decide, and engage with the world.

The crowdsourcing landscape is as diverse as its applications. It includes paid microtask platforms such as Amazon's Mechanical Turk\footnote{\url{https://www.mturk.com/}} and CrowdFlower (now Figure Eight),\footnote{\url{https://www.figure-eight.com/}. In the paper, we will continue to refer to CrowdFlower, as the work was carried on before the rebranding as Figure Eight.} alongside online labour marketplaces such as TaskRabbit\footnote{\url{https://www.taskrabbit.com/}} and UpWork,\footnote{\url{https://www.upwork.com/}} and open innovation contests in the style of Kaggle\footnote{\url{https://www.kaggle.com/}} and Innocentive.\footnote{\url{https://www.innocentive.com/}} It features volunteer citizen science systems such as Galaxy Zoo\footnote{\url{https://www.galaxyzoo.org/}} and SciStarter,\footnote{\url{https://scistarter.com/}} and games with a purpose (GWAPs) such as PhraseDetective\footnote{\url{http://anawiki.essex.ac.uk/phrasedetectives/}} and EyeWire.\footnote{\url{http://eyewire.org/explore}} Putting aside the principled differences between these forms of crowdsourcing, which make them from the offset amenable only to specific types of problems, the success of any crowdsourcing endeavour will depend on the ability of the `requester' - the person or institution reaching out to the crowd - to benefit from crowd outputs and attract a critical mass of contributors. The frame conditions can be challenging as well: to be effective, crowdsourcing needs to provide a real alternative to existing solutions of the problem the requester is trying to solve, in terms of quality, timeliness, and costs. Getting this mixture right is not trivial, as it requires insight into a wide array of subjects from artificial intelligence and data analysis to user experience design and behavioural sciences \cite{feyisetan2016please,feyisetan2017social}.

Aligning the motivations of the crowd with the goals of the requester and finding the right mix of rewards to drive purposeful crowd engagement are particularly critical. Existing crowdsourcing platforms already provide some level of support - whether paid microtasks, challenges, or citizen science projects, each of these forms of crowdsourcing makes some basic assumptions about what drives people to contribute to their `open call' \cite{howe2006rise}, and offers sometimes bespoke incentives to encourage a certain style of behaviour. For example, people register on Mechanical Turk primarily for financial reasons \cite{martin2014being,mason2010financial}, then for learning, and possibly for fun. However, this is sometimes predicated on the task paying well. On the other hand, in contests such as the Netflix prize\footnote{\url{http://netflixprize.com/}} or on TopCoder\footnote{\url{https://www.topcoder.com/}}, the crowd is mostly technology-savvy, and takes part mostly to gain a reputation, and then for the prize, or out of a desire to improve one's skills \cite{archak2010money}.  In citizen science, which is often devoid of financial rewards, the main motivator is a blend of learning, fun, and the wish to make a contribution to the world's knowledge \cite{raddick2009galaxy,tinati2017investigation}. Designing a successful crowdsourcing campaign may therefore require different incentives, which respond to the different reasons that drive people to engage. This is not easy to achieve, as layering incentives on top of crowd motivations to change behaviour has proven to engender unexpected effects - for example, studies have shown that paying for work completed out of intrinsic reasons may make people contribute less, a phenomenon referred to as `crowding out' or `over justification' \cite{frey2001motivation}.  Similarly,  incentivising people through gamification commonly fails to achieve significant, long-term results when there is a mismatch between the rewards induced through game elements and the needs, expectations and values of the crowd \cite{rughinis2013gamification,zichermann2011gamification}.

In our work we study the interplay between motivations and incentives in the context of \textit{paid microtask crowdsourcing}. As financial payments are just one, albeit an important, factor in using crowdsourcing effectively, we explore the use of points, badges, leaderboards and contents to boost task performance and crowd engagement. To do so, we use our own experimental crowdsourced-annotation platform \textit{Wordsmith}, which features paid microtasks at its core and allows requesters to layer different incentives on top of the financial remuneration and analyse their impact. 

An initial study of Wordsmith and the effects of gamified task design on paid crowdsourcing was presented in \cite{feyisetan2015improving}. It showed that designing a playful interface for annotation tasks - as opposed to the functional style common to most microtask platforms - encourages workers to engage with the task more, independently of the actual monetary reward. In the study, we were consistently able to achieve significantly lower costs ($5,708$ unique annotations collected via the game vs. $111$ unique annotations contributed in the control condition on CrowdFlower) and improve accuracy by almost $10\%$ compared to the baseline. Through furtherance incentives, which were offered to workers when we predicted they were about to leave, we could convince workers to continue annotating, generating up to five times more unique annotations at a comparable level of accuracy. These effects were stronger when the incentives were personalised, with $77\%$ of crowd workers deciding to complete one more image when confronted with a targeted furtherance incentives against $27\%$ in the experiments using a randomly selected one. A follow-up study looked at the effects of social incentives \cite{feyisetan2016please,feyisetan2017social} - we asked workers to complete annotation tasks in pairs and to encourage their co-worker to continue so that both parties receive their payments. Again, the results were very encouraging. Adding a collaborative component to the default microtask model led to a $233\%$ increase in task uptake, a $140\%$ increase in task output and a $6\%$ increase in inter-annotator agreement. We also demonstrated that social furtherance incentives create a win-win scenario for the requester and for the workers, by helping more workers get paid by re-engaging them before they drop out.

In this paper, we focus on improving another critical success factor of crowdsourcing, delivery time. We developed an extension of Wordsmith that allows crowd workers to pick up tasks in near-real-time and compete. Workers were recruited from CrowdFlower, but instead of carrying out their annotation work at their own pace, they were asked to pay attention to the time-sensitive element of the tasks, and aim for a faster turnaround. Central to the competition was the leaderboard, which decides which workers will receive a payment for their work: unlike in most paid microtasks projects, which pay workers per unit of work accepted by the requester, in our study, workers competed for top positions on the leaderboard to get paid. 

Crowd workers used Wordsmith to complete work for a fixed period of \textit{time}. They were rewarded only if they achieved a high enough position on the leaderboard - in other words, workers competed against each other and only a share of them (captured by a so-called \textit{reward spread}) received a payoff at the end of the contest. Leaderboard rankings were computed and updated on the fly as a function of the number of tasks completed and a heuristic approximation of their quality. The latter was based on previous work of ours in which we have studied different approaches to crowdsource the same annotation task, using standard paid microtask platforms \cite{feyisetan2015towards,feyisetanextended}.

The main intuition behind the contest model was that by designing the crowdsourcing exercise as a live contest, which is supposed to be completed in a relatively short period of time, we would create an environment in which results would be delivered quickly and with accuracy. In addition, as we did not pay all workers upfront or merely for being available, we could keep the overall costs lower than in other real-time crowdsourcing approaches, which often reward mere availability as opposed to work being done. In this context, we hypothesised that the number of workers rewarded impacted the output volume.

To test this, we ran experiments with three sets of reward spreads (top worker; top $5$ workers; and top $10$ workers). The experiments confirmed our initial assumption: the contest model resulted in more annotations at a rate that was twice as fast as the control condition reported in our previous work \cite{feyisetan2015towards}. Furthermore, increasing the reward spread led to an increase in task output. Rewarding more workers reduced the rate of worker attrition and kept more workers engaged. 

These results could help in finding the balance between theoretical guarantees with empirical evidence in order to select appropriate reward spreads, while scaling to data streams comparable to real-world situations.

\vspace{5mm}
\noindent \textbf{Summary of contributions}
Our platform Wordsmith offers a lens through which we can better observe the potentials of different incentive mechanisms in paid microtask crowdsourcing. Our main observations are the following: while \textit{financial payments} remain important, \textit{contests} can be extremely effective in encouraging workers to volunteer to carry out work at higher speed. \textit{Leaderboards} make workers more aware of their performance - they drop off when faced with a potential loss of utility, i.e., when they believe they might not be among the winners of the competitions and hence miss on the payments, but the faster-paced design does not have significant effects on output quality or user experience.  

The experiments shed light on possible design improvements of paid microtasks platforms to boost task performance, and make the overall experience more fair and rewarding for the workers. While we are not necessarily arguing for fully-fledged contest-based microtask platforms, considering specific gamification elements that are widely discussed in the literature \cite{kraut2012building,zichermann2011gamification} is worth further investigation. This is important not just for purely utilitarian motives on the side of the requesters, but also in the context of the ongoing debate on ethical and fair crowdsourcing \cite{irani2013turkopticon}. In the same time, competitions, which are widely used in other forms of crowdsourcing (for example, in citizen science \cite{tinati2017investigation} and in challenges \cite{tang2011reflecting}) offer interesting alternatives in situations when having a faster turnaround is important. 

Previous work has approached such questions mostly through studies of crowd motivations \cite{kaufmann2011more}, discussing the rich repertoire of extrinsic and intrinsic reasons that drive people to complete microtasks. Our experiments quantify some of their findings \cite{feyisetan2015towards,feyisetan2017social}. We deliberately chose a task that is well-known in the crowdsourcing literature \cite{gadiraju2014taxonomy}, as we were aiming for microtask designs which were only minimally influenced by interface or quality control aspects. For the same reasons, we opted for average market prices to reward participation; lower pays would have been less attractive (and unfair) for workers, higher ones might have appealed to people who were exclusively drawn to financially incentives. We believe more research needs to be done to build microtask platforms that are crowd-, as opposed to requester-centric, and reflect and support the values and motivations of the crowd as an integral part of their functionality. Our research gives evidence that such efforts could be beneficial for both workers and requesters - points, leaderboards and contests can be linked to intrinsic motivation. Workers not only completed more tasks than required, but some of them reported positively about the experiments on a community forum - for example, one post read "\emph{Hello everyone! lately I'm hooked on the multiplayer tasks, waiting for $100$ people to connect}", while another one claimed: "\emph{Hit the top $10$ today. I will hunt this problem again}".

Crowdsourcing, though riddled with its challenges from the technical to the ethical, is here to stay. Even as it morphs continually to the tune of academic and economic forces, it remains a force for good with groundbreaking discoveries being made daily directly and indirectly through the wisdom of the crowd \cite{good2011games}. Understanding the impact of incentives brings us one step closer to designing better, more responsible crowdsourcing systems. This is vital not just for research purposes, but also because of the influence crowdsourcing has on fields that have come to rely on it, including artificial intelligence, and those that draw inspiration from it, such as the so-called `gig' economy. 

\vspace{5mm}
\noindent \textbf{Previous publications} This paper builds upon previous work of ours, in particular on \cite{feyisetan2004semantics,feyisetan2015improving,feyisetan2015towards}. \cite{feyisetan2015improving,feyisetan2015towards} used earlier versions of Wordsmith to carry out their experiments. In \cite{feyisetan2015improving}, we presented a first version of Wordsmith and initial experiments into the effects of gamified task design on paid crowdsourcing. In \cite{feyisetan2004semantics,feyisetan2015towards}, we aimed to understand how to collect Twitter entity annotations via paid microtasks effectively. All works implemented a task model where workers complete tasks individually and are rewarded if they provide useful answers. In this work, we used the same Wordsmith platform and Twitter entity annotation task; however, we introduced contests as a new incentive layer on top of the base financial payouts.


\section{Background}
In this section, we briefly review some of the most relevant prior work pertaining to maximising the effectiveness of incentivised crowdsourcing. In particular, we focus on real-time crowdsourcing approaches that aim to generate results faster; and more generally, on methods that optimise some aspect of crowd performance, be that by offering bespoke incentives, by assigning tasks to those workers who are more likely to do them well and so on.  As much of this background literature is inspired by and explained using theories of human motivation and incentives design, we start with a synopsis of fundamental work in this space.

\subsection{Incentives in paid microtask crowdsourcing}
\subsubsection{Motivation theory}
Theoretical frameworks such as the Self Determination Theory (SDT) \cite{sdttheory} make the distinction between intrinsic and extrinsic motivation to study the reasons why individuals decide to contribute to a task. Intrinsic motivation covers tasks that are perceived as rewarding in themselves, and is seen as responsible for engagement in activities such as socialising and participating in volunteering projects \cite{tinati2017investigation}. On the other hand, extrinsic motives are related to factors that are not inherently related to the actual task, but are appealing for some external reason such as status or financial payments.

The SDT theory has been applied to motivation studies on paid microtask platforms. \cite{kaufmann2011more} mirrors the motivation types as \textit{fun} (intrinsic) and \textit{money} (extrinsic) in a survey of incentives on Mechanical Turk. Similarly, \cite{malone2010collective} describes incentives as either appealing to \textit{love}, \textit{glory}, or \textit{money}. More relevant to Wordsmith, SDT has also been applied extensively in gamification research \cite{deterding2011game,seaborn2015gamification}; collaboration and online communities \cite{raddick2009galaxy,nov2014scientists,tinati2017investigation}; and contests \cite{bennett2007netflix}.

\subsubsection{Monetary incentives}
Money is a natural incentive for carrying out work, hence paid microtasks remain one of the most prominent forms of crowdsourcing \cite{frei2009paid}. Research has shown that increased payment leads to faster task completion, but not to higher quality \cite{mason2010financial}. Several other factors play a role in obtaining useful task results: bonuses, worker perception, and the variation of payment sizes across tasks \cite{difallah2015dynamics,mason2010financial}. The drive to engineer optimal quality, speed and volume of work with minimal financial payments has made paid microtasks the subject of numerous discourses on the ethics of compensation \cite{irani2013turkopticon}. While monetary payments can be seen as an extrinsic motivator, other intrinsic factors have been known to contribute to sustained participation in paid crowdsourcing settings, including intellectual curiosity \cite{law2016curiosity}; the inherent nature of the task \cite{frei2009paid}; and interaction with other workers \cite{kittur2010crowdsourcing}. One of the questions we are trying to answer with Wordsmith is: \textit{How do we expand the currency of transactions on paid microtask platforms to transcend monetary payments, and encompass tasks that have an intrinsic appeal?}. We presuppose that framing tasks this way would be more attractive to workers, cost less for requesters, and make workers perform better. This led us to study other incentive mechanisms in our earlier works \cite{feyisetan2015towards,feyisetan2017social}, and the idea to leverage contests as a new incentive strategy.

\subsubsection{Gamified incentives}
Gamification is the use of game design in non-game contexts, with the aim to achieve the effects of fun and engagement that are created by playing a game \cite{zichermann2011gamification}. It often involves adding game-like rewards to non-game activities, include social elements to encourage teamplay or contests. We understand gamification in a broad context, comprising systems that build a complete game narrative around a task (e.g., FoldIt and EyeWire) \cite{good2011games}; employ tactics such as micro-diversions to fend off boredom \cite{dai2015and}; further a noble cause or stoke curiosity \cite{law2016curiosity}; or engineer game mechanics (such as points, badges, levels and leaderboards) into tasks \cite{deterding2011game}. In a work context, each of these classes of systems has been shown to improve productivity, drive engagement and effectiveness, and therefore reduce costs.  

Gamification practices have raised concerns about their potential to undermine innate intrinsic motivation via the crowding-out and over-justification effects discussed earlier - although this has been shown to be task-dependent \cite{mekler2013disassembling,seaborn2015gamification}. We also know that gamification can be more effective in collaborative settings; in other work of ours \cite{feyisetan2017social}, we designed a multiplayer Wordsmith extension that required consensus among pairs of paid crowd workers. This led to better outcomes when compared to the baseline, which was built around completing the same tasks individually \cite{feyisetan2017social}.  Similar findings were reported in the context of GWAPs \cite{von2008designing}, which often rely, unlike our work, on voluntary participation.

\subsubsection{Contest-based incentives}
Paid microtask crowdsourcing has traditionally been approached as an individualistic endeavour. Workers complete their tasks without depending on, or interacting with others. Other forms of crowdsourcing, such as volunteer citizen science, have experimented with models that encourage the crowd to discuss, exchange ideas, collaborate on tasks, or compete to improve their standing in the community \cite{soton419303,tinati2017investigation}.  In this context of paid microtasks, we seek to apply contest-based incentives to improve the delivery time of crowd responses. 

In order to carry out crowdsourcing in a contest setting, timely worker recruitment and task uptake are critical. This comes as a result of scenarios with near-real-time constraints - for example, when crowdsourcing is applied to fact-checking in news media, disaster management, or live data feeds. However, most solutions proposed in the literature are often too costly for large numbers of microtasks (see also Section \ref{sec:contests} below for an overview of near-real-time crowdsourcing research). Contests recruit crowd workers who are attracted to the thrill of `glory' and competition to carry out tasks quickly. This model has been seen on software coding platforms such as TopCoder \cite{archak2010money}, where over a million workers vie to complete challenges, or in citizen science \cite{soton419303}. Contests leverage on the urgency that comes from a fixed completion time-frame, and the satisfaction brought by a sense of winning. They are primarily deployed when the requester seeks one best, or final answer (as opposed to an aggregation of worker results, which is more typical in microtask crowdsourcing), or to increase participation. For example, the Netflix \$1 million dollars challenge to build a better recommendation algorithm \cite{bennett2007netflix} falls in this `single best result' category. EyeWire, a game with a purpose which creates a visual map of the human brain, routinely uses different kinds of challenges to increase the volume of tasks completes and keep the crowd engaged \cite{soton419303}. Remuneration in contests range from a winner-takes-it-all model, which compensates only the best participant, to more relaxed models that pay contributors who make submissions above a certain threshold, with several theoretical frameworks set up to determine the optimal allocation of prizes \cite{moldovanu2001optimal}. In this paper, we designed our contest to cover tasks that required an aggregation of results (as opposed to a single best response), while obtaining responses as quickly as possible. This type of approach is relevant in domains such as disaster relief and real-time visual aids for blind people \cite{bigham2010vizwiz}.

The literature on contests goes well beyond the focus of our work, which is paid microtasks. The history of contests dates back to as far as $1714$, when the British Parliament ran one to determine the longitude at sea to within half a degree \cite{moldovanu2001optimal}. Several theoretical studies have looked at optimal contest design \cite{chawla2015optimal} and the optimal allocation of prizes \cite{moldovanu2001optimal}. 
A survey of experimental research of contests is available from \cite{dechenaux2015survey}. Researchers have investigated how contents unfold on existing platforms such TaskCN \cite{liu2011crowdsourcing},  TopCoder \cite{archak2010money} or EyeWire \cite{soton419303}. Empirical studies have shown the effects of increased payoff as an indicator of contestant performance. Further on, \cite{boudreau2011incentives} considered over $9,000$ contests hosted on the same platform in order to understand the effect of participant numbers on the performance of individuals. A second group of empirical work focused on bespoke experimental setups for crowdsourcing. For example, \cite{rokicki2014competitive} looked at the effect of varying monetary schemes and information policies in individual contests, while \cite{rokicki2015groupsourcing} explored the same problem alongside team formation strategies. Both papers bear similarities to our scenario, in which contestants strategically decide whether they continue to take on more tasks or leave Wordsmith. However, they do not propose ways to predict worker exit to optimise task completion and delivery. \cite{norrander2006attrition} introduced a duration model, which we also attempt to analyse in our work, which shows the length of candidacies and factors associated with candidate exits. 

Finally, our work builds on literature which studies the war of attrition. In contests, each participant enters knowing their own skill and costs, but not those of the other contenders. Participants consider dropping out when they learn their opponents' strengths and discover that staying would be unprofitable. A theory of how this phenomenon operates in duopolies was presented by \cite{fudenberg1986theory}. \cite{krishna1997analysis} presented its relation to an all-pay auction while, while \cite{norrander2006attrition} discussed attrition and exit in political primaries. \cite{moldovanu2012carrots} looked at contests with exits where contestants have the option to drop out or not to participate with the introduction of costless punishments. Since our objective was to maximise the total utility generated in real-time, we did not use punishments, which some early pilots we carried out revealed to increase the attrition rate.

\subsection{Making microtask crowdsourcing more effective}
Several descriptive frameworks have been proposed in the literature to capture the nuances of incentives engineering beyond simplistic \textit{'fun or money'} considerations. Some of these include \textit{MICE} (Money, Ideology, Coercion, Excitement) \cite{randyburkett2013}; \textit{RASCLS} (Reciprocation, Authority, Scarcity, Commitment, Liking, Social Proof) \cite{randyburkett2013}; and \textit{SAPS} (Status, Access, Power and Stuff) \cite{zichermann2011gamification}. The latter is intended to represent a system of incentives from the most desired to the least desired, and from the cheapest to the most expensive. We adopted this framework in Wordsmith.

Mechanisms for the effective allocation of incentives have been studied in market and auction platforms, wireless and peer-to-peer networks and corporate organisations. In the context of crowdsourcing, several studies have been carried out that apply \textit{game-theory} techniques to incentive design \cite{xie2014fair,yang2012game}. These papers focus on financial incentives and a premise of inter-player strategy dependency. Not all crowdsourcing tasks can be modelled in this way; we adopt a probabilistic approach based on prior player behaviours to predict appropriate incentives beyond the purely financial (see also \cite{feyisetan2015towards}). Similar techniques are used for various purposes in crowdsourcing design, in particular to inform the assignment of tasks to workers or to predict task completion \cite{demartini2013large,sheng2008get}.

A large body of work has been dedicated to task and workflow design, as well as quality control (see, for instance, \cite{michelucci2013handbook} for a good compilation). We take their findings into account when implementing the basic CrowdFlower interfaces, checking for spam and validating and aggregating the results.

\subsection{Improving delivery time}
\label{sec:contests}
Longitudinal studies of crowdsourcing marketplaces such as Mechanical Turk \cite{difallah2015dynamics} reveal how microtasks have evolved to support scenarios in which work cannot take days or hours, but must be delivered in seconds and under \cite{bernstein2011crowds}. For near-real-time behaviour, two components are critical: prompt recruitment of workers and fast completion of tasks. In this section, we compare our approach with previous research in microtask crowdsourcing, which has proposed improvements to these components.  

\subsubsection{Timely worker recruitment} Several models have been put forward to ensure timely availability of crowd workers. For example, Adrenaline \cite{bernstein2011crowds} employed a retainer model, while in Viz-Wiz \cite{bigham2010vizwiz} workers signed up in advance and were kept engaged, at a cost, until the work arrived. Other approaches relied on very large crowds \cite{lasecki2014architecting}, using queuing theory or predictive recruitment to model workers' arrival and task assignment \cite{bernstein2012analytic}, or repeatedly posting the tasks to make them more visible to workers \cite{bigham2010vizwiz}. In our experiments, we incorporated a combination of these ideas to generate our on-demand crowd: we repeatedly posted tasks, recruited in advance, requested for a larger crowd than required, and used an audio alert to prompt workers when the contest started. At the same time, by virtue of our contest model, we rewarded only a small share of the workers, hence keeping the costs down.

\subsubsection{Timely task completion} Some of the techniques found in the related literature include: \emph{rapid refinement}, which dynamically narrows the search space following signs of worker agreement \cite{bernstein2011crowds}; \emph{stream parallelism}, which splits the task into sub-tasks and assigns them to different pools of workers in parallel \cite{bigham2010vizwiz}; \textit{temporal division}, which breaks the task  down into time slices as more work becomes available \cite{lasecki2011real}; \textit{warping time} \cite{lasecki2013warping}, which allows workers to listen to audio streams at reduced speeds; and a \textit{Map Reduce} alike to organise complex workflows \cite{kittur2011crowdforge}. Our model uses stream parallelism to assign work to several groups of workers, and variants of temporal division and time warping to display tasks to each overlapping batch of workers for an extended period of time. Combined with the mixed cardinal-ordinal contest, these techniques create a powerful mechanism to solve tasks twice as fast as the evaluation baseline (see Section \ref{sec:results}).

\section{Contest-based annotation task}
In our experiments, we used a text annotation task model in relation to real-time crowdsourcing. For this reason, the task consists of a total of \textit{n} posts, \begin{math}P = \{p_1, ..., p_n\}\end{math}, each containing \textit{m} entities \begin{math}E = \{e_1, ..., e_m\}\end{math} to be annotated, where \begin{math}m < M_i! + M_i\end{math} and \begin{math}M_i\end{math} is equal to the number of text tokens in post \begin{math}p_i\end{math}. The posts arrive at a constant rate \begin{math}\lambda\end{math} and each has a processing rate of \begin{math}\mu\end{math}. There are \textit{n} workers in a pool to serve the task queue such that, to keep up with the requests, the ingress load (task intensity) \begin{math}L = \lambda/\mu\end{math} stays less than the number of workers \textit{n}, i.e., \begin{math}L < \textit{n}\end{math}. 

Tasks that are not solved are dropped of the queue as opposed to being kept indefinitely in the buffer \cite{bernstein2012analytic}. They are solved using a first-in-first-out scheduling and processing scheme, in which already recruited workers are sought to carry out new tasks (as opposed to recruiting additional workers for new tasks). Therefore, the requester is looking for an optimal processing rate \begin{math}\mu\end{math}, and needs to keep workers motivated to carry out as many tasks as possible. 



\vspace{5mm}
\noindent \textbf{Requester} The requester asks the crowd to complete a series of tasks in real-time. The requester needs to determine the experiment setup in terms of: (i) the number of contestants; (ii) the number and size of prizes; and (iii) the contest constrains to maximise the effort exerted by all contestants \begin{math}\sum^{w}_{i = 1}\epsilon_i\end{math}. This is different from contests such as the Netflix Challenge \cite{bennett2007netflix}, where the principal's objective was to elicit a single best response to a task.

The requester does not only desire to maximise the total exerted effort, but also to maximise some utility function of output quality \begin{math}\sum^{w}_{i = 1}q_i\end{math}. The requester therefore needs to maintain incentives for highly skilled workers, and motivate low skilled workers to exert more effort while adjusting the prize spread. 

The requester defines: (i) a completion time constraint \textit{T}, which depends on the number of posts \textit{n} and their arrival rate \begin{math}\lambda\end{math}; and (ii) a quality constraint \textit{Q}, which denotes the minimum number of labels expected from each worker to be eligible for payment. The latter is essential in hybrid tasks; for example, the task might have been pre-labelled by a machine to determine the probable number of named entities (this serves as the quality constraint \textit{Q}), while the crowd workers identify those entities and type them \cite{feyisetan2015towards}.

The requester is able to observe the baseline quality of each worker's output, based on the pre-computed number of entities and the number of entities submitted by the worker. The requester is then able to use this information to construct the contest by assigning a quality score to every crowd answer - therefore, contestants are ranked not only based on their effort (number of posts annotated), but also on the quality of their output. The reward mechanism awards a prize \begin{math}A_j\end{math} to a worker \begin{math}w_j\end{math} within a reward spread (e.g., the worker was within the top $5$ or top $10$).

\vspace{5mm}
\noindent \textbf{Workers} There is a set of \textit{n} workers, \begin{math}W = \{w_1, ..., w_n\}\end{math} participating in the contest, each with the ability to carry out entity annotations. Each worker \begin{math}w \in W\end{math} has a private skill level \begin{math}\varsigma_i\end{math} (also known as expertise or ability), and for each post in an annotation task, chooses to exert a level of effort \begin{math}\epsilon_i \geq 0\end{math}. The skill level is drawn independently of other workers from the interval \begin{math}\varsigma_i \sim [0,1], \forall w \in W\end{math}, according to a distribution function \textit{F} with density \begin{math}f(\varsigma) = dF(\varsigma) > 0\end{math}. 

The effort exerted is drawn from the interval \begin{math}\epsilon_i \sim [0,\epsilon]\end{math}, in which the maximum effort expendable is constrained by the running time \textit{t} of the contest, which in turn is a function of posts per unit time and total number of posts. The quality \begin{math}q_i\end{math} of each worker \begin{math}w_i\end{math} is determined by: (i) the skill level \begin{math}\varsigma_i\end{math}; (ii) the effort exerted \begin{math}\epsilon_i\end{math}; and (iii) a requester variable \begin{math}\delta_i\end{math}. The requester variable \begin{math}\delta_i\end{math} is a function of the requester's review process and perception of quality, in comparison with the worker's internal tagging bias, which is markedly present in human judgement tasks. This value is constant across annotation posts \begin{math}\forall p_i \in P\end{math}; therefore, two workers exerting the same effort to annotate the same post would differ only on their skill, since the requester's variable is constant for that post. The quality of a submission is thus given as: \begin{math}q_i = \varsigma_i\epsilon_i + \delta_i\end{math}. In our experiments, the requester's variable was a measure of results in a pre-computed gold standard set \cite{feyisetan2015towards}.

Each worker \begin{math}w_i\end{math} seeks to maximise their expected utility. This depends on the number and value of prizes, and on the number of contestants and value of their efforts. A worker's utility \begin{math}U_i\end{math} is given by \begin{math}U_i = V_j - c(\epsilon_i)\end{math}, if the worker \begin{math}w_i\end{math} wins prize \begin{math}V_j\end{math}, or by \begin{math}U_i = - c(\epsilon_i)\end{math} otherwise, where \begin{math}V_i\end{math} is one of \textit{k} prizes to be awarded by the requester, and \begin{math}c(\epsilon_i)\end{math} is the worker's cost function, which is a strictly increasing function dependent on exerted effort where \begin{math}c(0) = 0\end{math}. Each prize \begin{math}V_j\end{math} above point \textit{k} is positive, or zero otherwise. We did not model negative rewards (punishments), as they did not seem to have the desired effects in early pilots we carried out.

During the contest, crowd workers are shown a list of tasks which arrive in near-real-time (in our experiments we used Twitter posts, but they could equally be images as in \cite{bigham2010vizwiz} or other objects). The workers are to annotate as many of them as possible within a time window. The incentives mechanism awards a prize \begin{math}A_j\end{math} to a worker \begin{math}w_j\end{math} within a reward spread (for example, the worker is within the top $5$ or top $10$).

\vspace{5mm}
\noindent \textbf{Exit} A worker would always seek to maximise their expected utility given the number and value of prizes; the number of contestants; and the value of their efforts. In our experiments, it was possible for workers to view their ranked position in real-time with respect to their closest contenders using a $k$-neighbours leaderboard, as presented in the medium information policy contest strategy by \cite{rokicki2014competitive}. A worker far outside the reward spread might inadvertently decide to exit the contest to avoid further loss of utility. The worker close to the reward spread might, however, decide to remain in the contest in the hope of displacing a close contender. 



\section{Contests as continuous time Markov chains}
\label{sec:ctmc}
\subsection{Model}
We modelled the contest-based annotation task as a continuous time Markov chain (CTMC). To understand CTMCs, we first define some key concepts. 

A \textit{stochastic process} is defined as a collection of random variables. It describes the evolution in time of a random phenomenon. Therefore, we say that the stochastic process is indexed by time i.e., the random variables $X$ for each time point $t$ is given by $X=\{X(t):t \in T\}$. When $T = \mathbb{N}$, it is a discrete-time process, however, when $T = [0,\infty)$, it is a continuous time process.

The \textit{Markov property} refers to the \textit{memoryless} property of a stochastic process. A stochastic process $X(t)$ has the Markov property if the conditional probability distribution of its future evolution depends only on its current position, not on how it got there. i.e. given ${X(t) : t \geq 0}$ with state
space $S$, we say it has the Markov property if:

\begin{equation}
P(X(t) = j|X(s) = i, X(t_{n-1}) = i_{n-1},...,X(t_1) = i_1) = P(X(t) = j|X(s) = i),
\end{equation}

where $0 \leq t_1 \leq t_2 \leq ... \leq t_{n-1} \leq s \leq t$ is a sequence of $n + 1$ times and $i_1, i_2,...,i_{n-1}, j \in S$ are states in the state space. Therefore, a continuous time stochastic process $X=\{X(t):t \geq 0\}$ is called a continuous-time Markov chain if it exhibits the memoryless Markov property.

Revisiting our contest-based annotation task, we assume that the $n$ Twitter posts arrive according to a Poisson process with rate $\mu$. This Poisson process is a continuous-time Markov chain with state space $S = \{0, 1, 2, ...\}$ representing the worker annotations. Therefore, at any time $t$, a worker $i$ is in a state of annotating the $j^{th}$ tweet. The time it takes for the worker $i$ to annotate this tweet is represented as an exponential random variable, which transitions the worker to state $j + 1$.

The question then is: what `explains' the rate at which a worker annotates each tweet. In our modelling in this section, and in subsequent empirical experiments, we evaluate one variable to explain the annotation rate. We have earlier described this as the \textit{reward spread} which governs, at any time $t$, if worker $i$ is \textit{currently} eligible to receive a payment based on their leaderboard rank.

To model our contest-based annotation task as a continuous time Markov chain, we define the following variables:

\noindent \textbf{Observed variables}:
\begin{itemize}
    \item $S_i(t)$: the number of annotations of worker $i$ in finite time $t$
    \item $t_{j}^{i}$: the time of the $j^{th}$ annotation of worker $i$
    \item $\tau_{j}^{i}$: the annotation time for worker $i$ after the $j^{th}$ annotation
\end{itemize}

\noindent \textbf{Hidden variable(s)}:
\begin{itemize}
    \item $\theta_i$: the annotation rate parameter for worker $i$
\end{itemize}

\begin{figure}[h]
\centering
\includegraphics[width=0.60\columnwidth]{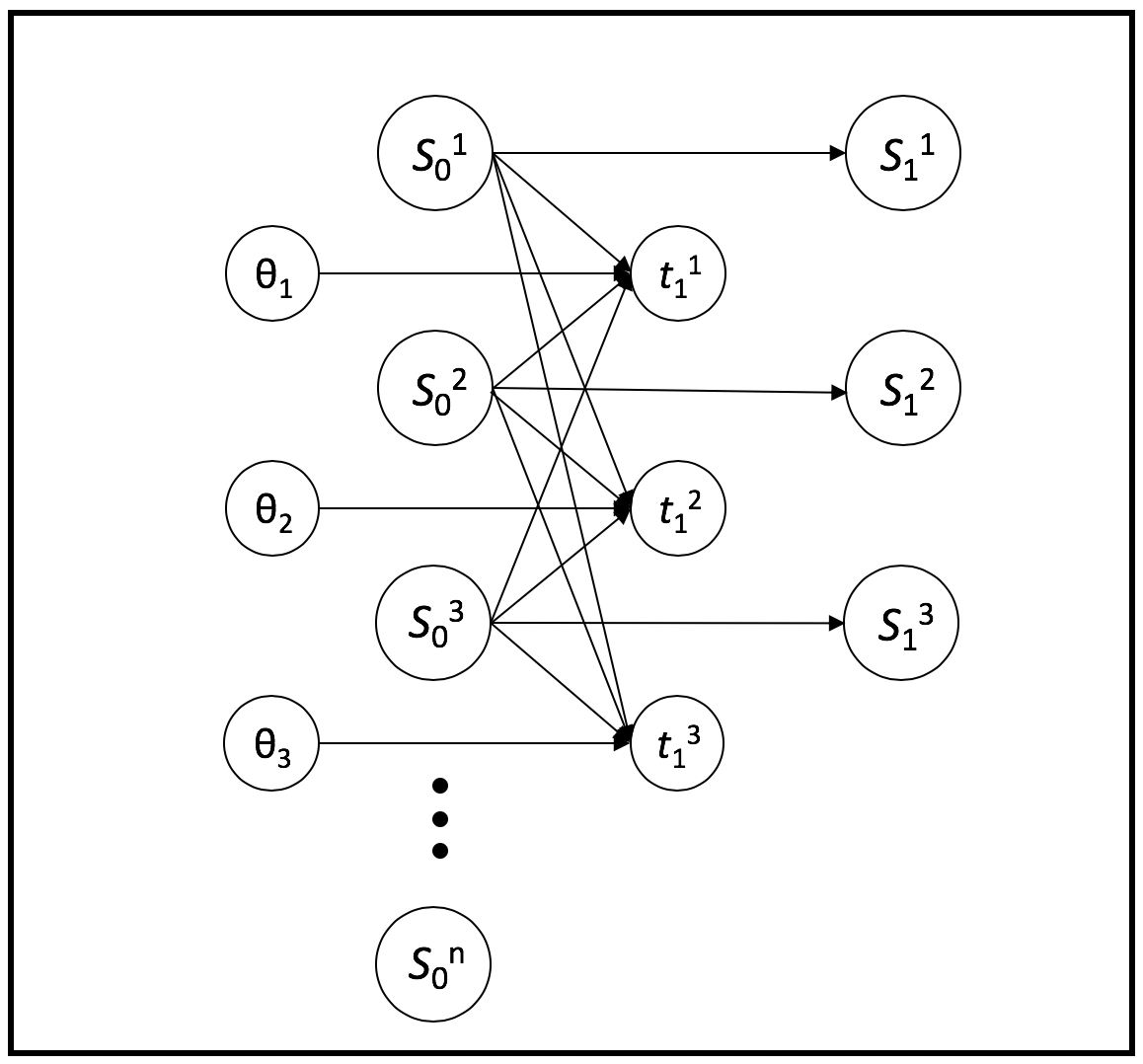}
\caption{State transitions}
\label{fig:transitions}
\end{figure}

The time a worker spends on annotating a tweet is calculated as:

\begin{align*}
\tau_{j}^{i} \sim P\Big(\tau^{i} \;|\; \{S_{k}(t_{k}^{i})\}_{k};\; \theta_i\Big)
&=
\exp\Big(\tau^{i};\; \lambda \;.\; f(\{S_{k}(t_{k}^{i})\}_{k};\; \theta_i)\Big) \\
&= f\Big(\{S_{k}(t_{k}^{i})\}_{k};\; \theta_i\Big) \exp\Big(f(\{S_{k}(t_{k}^{i})\}_{k};\; \theta_i)\Big)\; \tau_{j}^{i}\\
\end{align*}

Furthermore, the time of the next annotation, and the next state can be expressed as:
\[
t_{j+1} = t_{j} + \tau_{j}^{i};\; S(t_{j+1}) = S(t_{j}) + 1\\
\]

The parameter $\lambda$ describes the annotator `behaviour' within the \textit{reward spread}. This biases workers to either exert more or less effort based on their rank i.e., if a worker is currently eligible to be paid, they might either choose to exert more effort to extend their lead, or they might relax since they are already in the top ranks. Similarly, a worker outside the reward spread might either give up or exert more effort to displace one of the top workers. We modelled $\lambda$ using the gamma distribution $f(x; \alpha, \beta)$, and half-normal distributions $f_{Y}(y; \sigma)$, where $Y = |X|$ and $X \sim \mathcal{N}(\mu,\,\sigma^{2})$. The half-normal distribution biases our belief that some workers will increase their effort, especially when they are outside but close to the reward spread.

\[
    \lambda \; \sim \; 
\begin{dcases}
    \frac{\beta^{\alpha}x^{\alpha -1}e^{-\beta x}}{\Gamma(\alpha)},& \text{if worker is eligible to be paid}\\
    f(x; \alpha, \beta) \;+\; \frac{\sqrt{2}}{\sigma \sqrt{\pi}} \exp \Big(-\frac{y^2}{2\sigma^2}\Big),              & \text{otherwise}
\end{dcases}
\]

Therefore, simplifying and expressing this for a single worker:
\begin{align*}
P\Big(\{t_{j}^{i}\}_{j=0}^{N_i}\; \{S_{i}(t_{j}^{i})\}_{j=0}^{N_i} \;|\; \theta_i\Big)
=
\prod\limits_{j=0}^{N_i} f\Big(\{S_{k}(t_{j}^{i})\}_{k};\; \theta_i\Big) \exp\Big(f(\{S_{k}(t_{j}^{i})\}_{k};\; \theta_i)\Big)\; \tau_{j}^{i}\\
\end{align*}

Computing this for all the workers, we get:
\begin{align*}
P\Big(T, S \;|\; \theta_i\Big)
=
\prod\limits_{i=1}^{n} \prod\limits_{j=0}^{N_i} f\Big(\{S_{k}(t_{j}^{i})\}_{k};\; \theta_i\Big) \exp\Big(f(\{S_{k}(t_{j}^{i})\}_{k};\; \theta_i)\Big)\; \tau_{j}^{i}\\
\end{align*}

Expressing this as a log loss for training a learning model, we obtain:
\begin{equation} \label{eq:1}
- \log P\Big(T, S \;|\; \theta_i\Big)
=
\sum\limits_{i=1}^{n} \sum\limits_{j=0}^{N_i} - \log f\Big(\{S_{k}(t_{j}^{i})\}_{k};\; \theta_i\Big) + f\Big(\{S_{k}(t_{j}^{i})\}_{k};\; \theta_{i}\Big); \tau_{j}^{i}\\
\end{equation}

\begin{figure}[h]
\centering
\includegraphics[width=0.80\columnwidth]{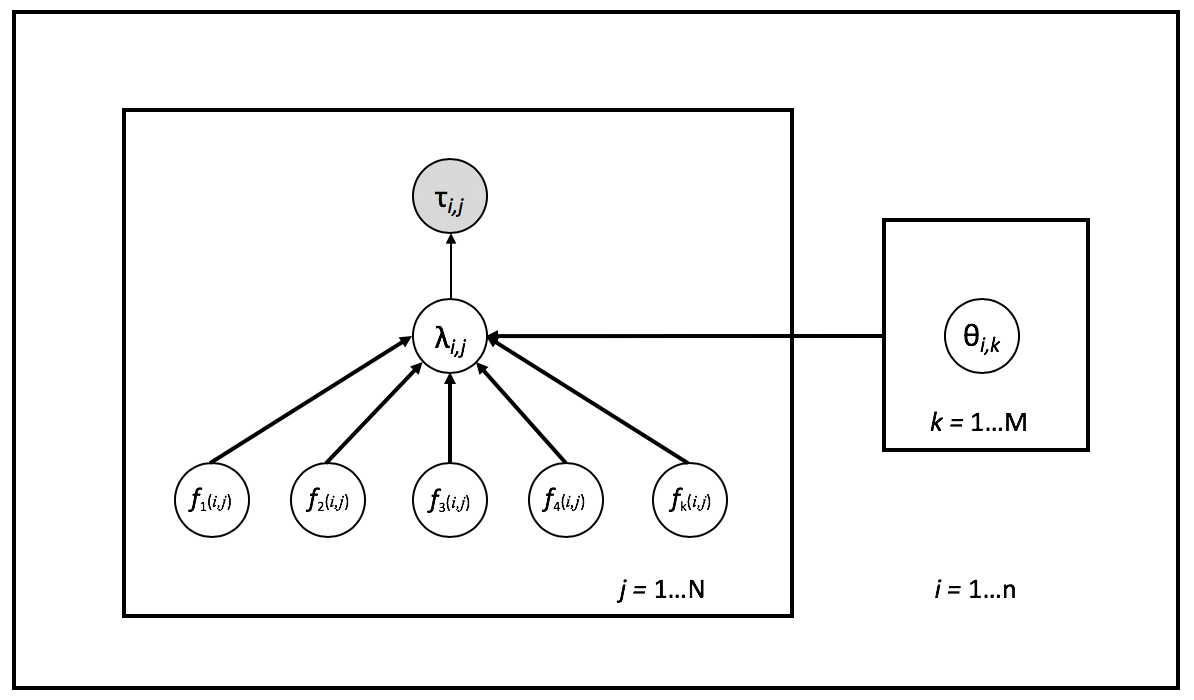}
\caption{Plate notation for CTMC contest-based annotation task}
\label{fig:pgm}
\end{figure}

Fig. \ref{fig:pgm} represents a graphical model that presents the variables $\tau$, $\lambda$, $f$ and $\theta$ in our model. The arrows describe how each variable influences or affects another. Therefore, obtaining the values $\tau_{j}^{i}$ (i.e., for each worker $i$ at every annotation point $j$) helps us predict how much time the annotator will spend in the entire contest.

\subsection{CTMC simulations}
We ran a series of simulations to test the CTMC model for contest annotations tasks. The independent variable is the reward spread which we hypothesise can explain the annotation rate of workers in the contest.

To aid visualisation, we ran the simulation using $5$ annotators, $100$ annotations and $200$ time steps. First, we generated $\lambda_i$ values which represent the behaviour of each of the $5$ workers when they are within the payment or no payment leaderboard ranks.

\begin{figure}[h]
\centering
\includegraphics[width=0.70\columnwidth]{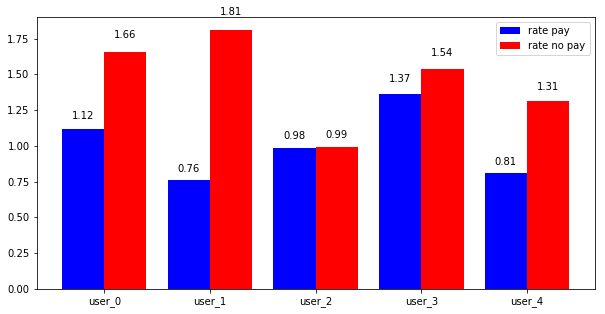}
\caption{$\lambda$ rates for each worker representing how much effort a worker exerts when they are within and outside the reward spread}
\label{fig:reward_spread_1}
\end{figure}

Fig. \ref{fig:reward_spread_1} displays randomly generated $\lambda_i$ values for each worker$_i$. From Fig. \ref{fig:pgm}, we know $\lambda_i$ influences $\tau_{j}^{i}$, therefore, we can simulate a contest run to see how the worker's strategies pan out: worker$_1$ who exerts significantly more effort only when they are not getting paid, or worker$_3$ that exerts much effort even when they are in the lead.

\begin{figure}[h]
\centering
\includegraphics[width=0.99\columnwidth]{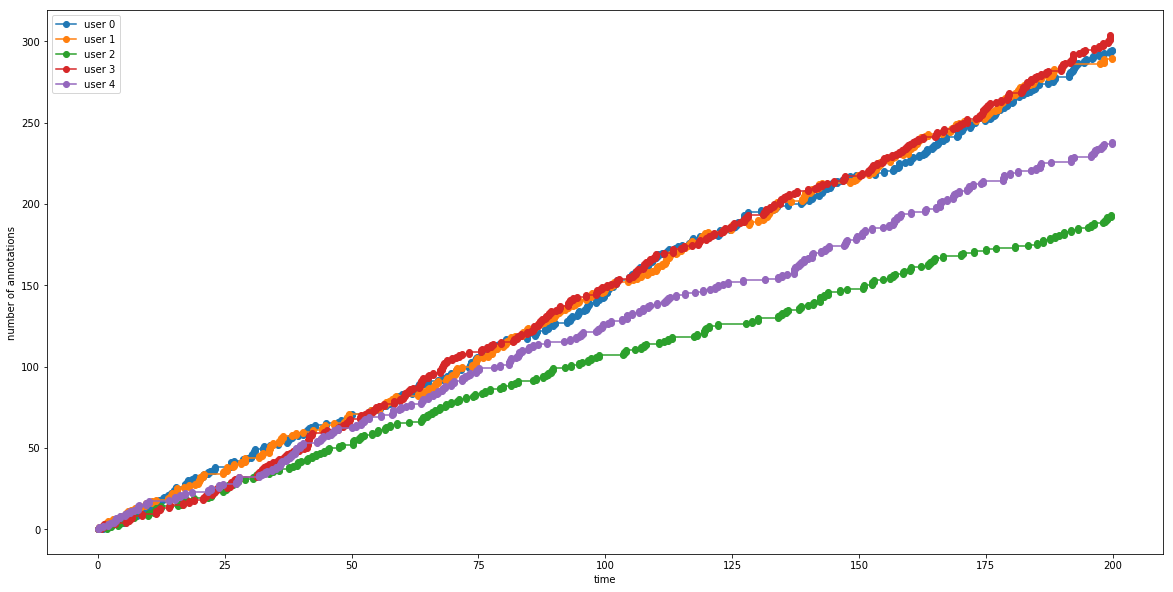}
\caption{Contest simulation using generated $\lambda$ rates}
\label{fig:contest_run}
\end{figure}

The results in Fig. \ref{fig:contest_run} paint a picture of how the change in annotation rates within and outside the reward spread affect how the entire contest turns out. This becomes more evident as the race progresses, in this case, with worker$_3$ coming out at the top with their $\lambda$ annotation strategy.

Therefore, Fig. \ref{fig:contest_run} illustrates that we can go from defining values for $\lambda$ to simulating an entire contest run. The challenge then is to do the reverse, that is, to infer the values of $\lambda$ from observing the transition $\tau$s in a contest. Learning this would allow us to modify the reward spread in a real-life contest, which would in turn modify $\lambda$, resulting in a desired contest outcome. The desired outcome from a requester's point of view would be to maximise the total annotations from all workers.

Using the loss function derived in Eq. \ref{eq:1}, we trained a model to learn the $\lambda$ rates of each of our $5$ workers. We used the worker's rank at each time step, the elapsed time, the number of annotations left, and whether the worker was within or outside the reward spread as features to train the model.

\begin{figure}[h]
\centering
\includegraphics[width=0.99\columnwidth]{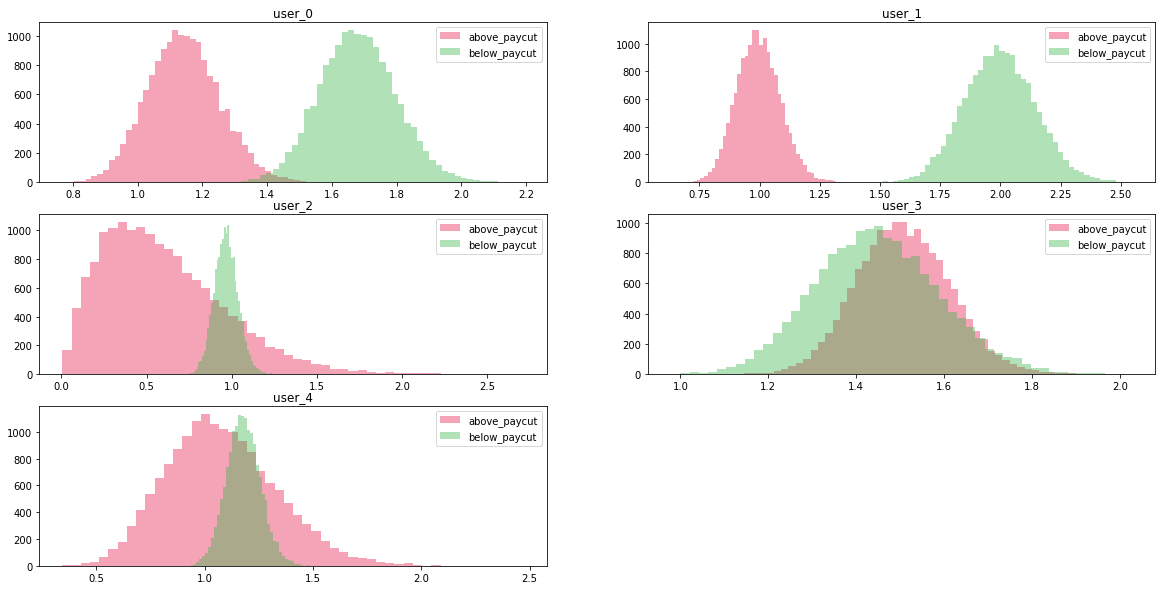}
\caption{Learned values of $\lambda$ rates for each worker}
\label{fig:model_infer}
\end{figure}

The results in Fig. \ref{fig:model_infer} illustrate that we can learn the $\lambda$ rates of workers which describe how they perform when they are within and outside the reward spreads. For example, observing the results of worker$_0$ in Fig. \ref{fig:model_infer} shows that the $\lambda_0$ rate when the worker is outside the reward spread is normally distributed around a mean value of $1.15$ which is very close to the value of $1.12$ generated and displayed in Fig. \ref{fig:reward_spread_1}. Similarly, the $\lambda_0$ rate when worker$_0$ is within the reward spread is normally distributed around $1.7$ which is close to the generated value of $1.66$.

\section{The Wordsmith platform}
\label{sec:platform}
Having laid the theoretical models underpinning our experiments, we now give a general overview of the Wordsmith platform and elaborate on those components that are relevant to our contest experiments. 

\subsection{Overview}
\begin{figure}[ht]
\centering
\includegraphics[width=0.95\columnwidth]{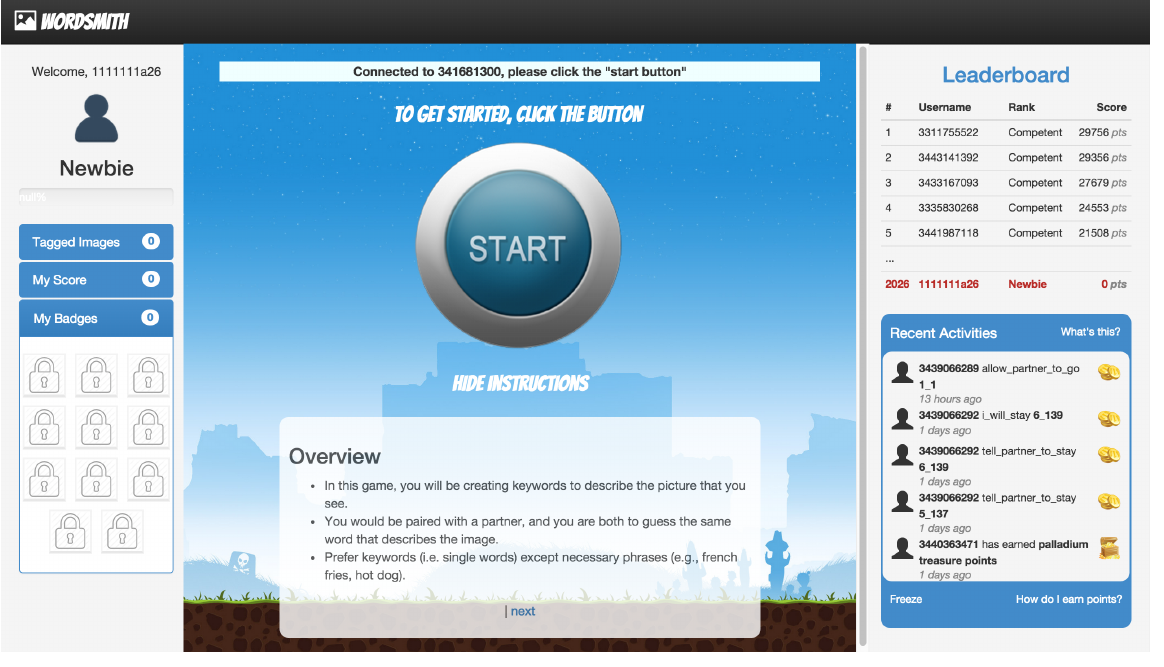}
\caption{Wordsmith start interface. On the left hand side, the worker sees their current level, the number of tasks completed, the number of points obtained, and their badges. On the right hand side, they can consult the leaderboard and a feed of activities of other workers who engaged with the task.}
\label{fig:wordscmithgui}
\end{figure}

Wordsmith is a platform for carrying out paid microtask crowdsourcing; Fig. \ref{fig:wordscmithgui} depicts a screenshot. It supports a range of annotation tasks, including text and image annotation. Further on, Wordsmith features multiple task, feedback and interface elements that allow for single-worker, multi-worker and contest-based task completion. 

The general task annotation works as follows: Workers are recruited from an existing microtask platform, for instance CrowdFlower, and redirected to our platform; the entry on CrowdFlower explains the task (e.g. identify people in tweets); the requirements (e.g. annotate at least ten images); and the financial remuneration. It also includes a a tutorial consisting of screenshots of the task interface on Wordsmith to give workers a step-wise idea on how to approach the task effectively. 

Requesters specify the conditions for workers to receive their payments -- for instance, a minimum number of annotations being carried out. However, workers are generally allowed to carry out as many tasks as they wish, beyond the task threshold which guarantees their payment. The requester can hence take advantage of the fun factor and game-like immersion that comes from completing the task in a gamified setting. Workers are also allowed to skip tasks - they can still receive their reward as long as overall they meet the conditions of the requester.

Requesters can also define game levels associated with an increasing volume of work to be completed or, if possible, with work of increasing difficulty. They also determine which worker activities are rewarded with gamification elements, including points, badges, and, as in \cite{feyisetan2015towards}, additional payments. 



\subsection{Contest design}
\label{sec:contestdesign}
The contest setting on Wordsmith allowed for multiple workers to connect to the system at once and carry out annotations simultaneously.

The current implementation of contests in Wordsmith builds upon the crowdsourced NER work which was presented in \cite{feyisetan2015towards,feyisetanextended}. In that work, we discussed the basic design of a text annotation microtask for Twitter posts, while here the focus is on the specific challenges arising from the fact that the annotations need to be delivered under significant time constraints.  While the contest model was tested for NER on microposts, it can be used for other types of labelling tasks that arrive in a real-time feed, with or without predefined label categories, for example for images like in \cite{bigham2010vizwiz}. For other text streams, the system would need first to divide the text into smaller chunks, which can still be treated by workers as microtasks in terms of the length of the text. \cite{bernstein2015soylent} offers interesting insights into how to process larger text documents via paid microtask crowdsourcing.

\subsubsection{Input and output} The system takes in a raw input of streaming posts and performs an initial sequence of processing on it. For our study, we focus on filtering out non-English tweets using the language tag of the incoming tweets. We then carry out part-of-speech tagging to recognise tweets with proper nouns. This is used to build a pseudo-quality score for each annotation (presented earlier as the requester's variable \begin{math}\delta_i\end{math}) i.e., if our POS tagger detects two different contiguous proper noun sets, we can expect an annotation result of at least two entities (although this does not hold strictly if there are no proper nouns e.g., entities might be recognised as noun phrases by some taggers).

The system outputs a processed stream of English tweets, \textit{(tweet1, ..., tweetN)}, where each tweet is represented as a tuple containing a reference identifier, the tweet string and the associated requester's variable \begin{math}\delta_i\end{math}. Each tweet in the stream advances in linear discrete time at a constant rate, with each time point represented as unique integer value in seconds (although, worker annotation and exit is represented in milliseconds). 

\subsubsection{Temporal division and stream parallelism} The stream is bucketed into distinct time intervals using \textit{windows}. A window consists of a constant number of microtasks emitted per unit time, which may or may not have been built over a buffer depending on varying throughput levels. Within each window task slice, microtasks are clustered and parallelized to different workers. Each cluster is a task unit consisting of a list of objects which is allocated to each worker for annotation. This follows a Map Reduce paradigm wherein each worker has a small task unit to solve, which is recursively built up to the final solution for the requester. The Map process involves local processing on individual nodes (annotating workers), while the Reduce process involves the merging of results to select best responses for overlapping task annotations.

\subsubsection{Contest interface}
\begin{figure}[ht]
\centering
\includegraphics[width=0.95\columnwidth]{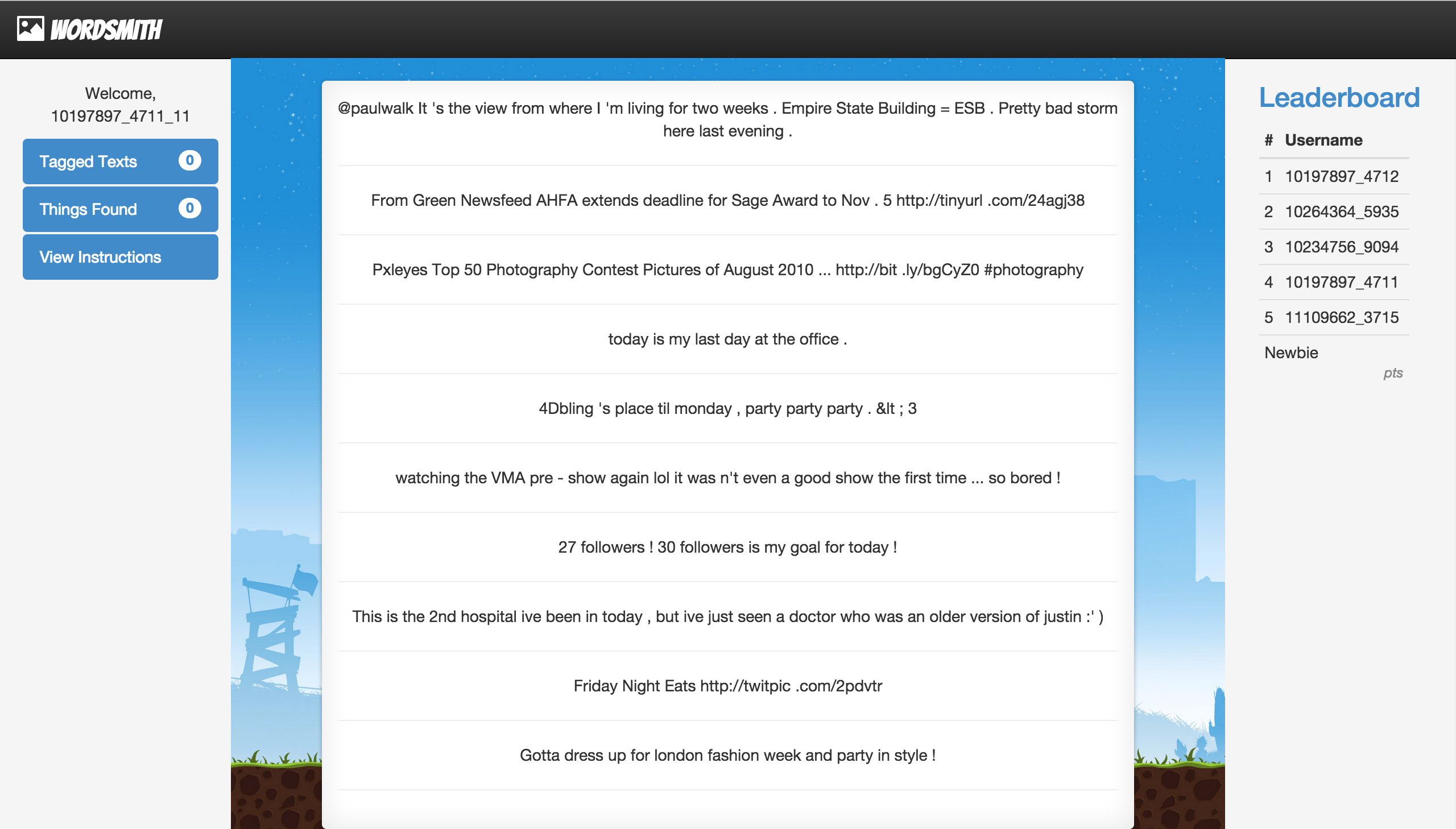}
\caption{Wordsmith contest interface. The central pane contains a microtask feed, which is updated according to the parameters fixed by the requester. Each task requires the player to identify instances of people, locations and organisations. On the right hand side, the players can see how they perform compared to others.}
\label{fig:exp3_contest}
\end{figure}

The contest interface consisted of a central annotation panel, in which the workers saw the current list of text snippets as described in \cite{feyisetan2015towards,feyisetanextended}. After selecting the piece of text to analyse, the workers could either annotate the entities using predefined categories, mark the post as having no entities, skip the task all together, or go back to the list. Fig. \ref{fig:exp3_contest} shows a screenshot.

Each worker received a baseline score $x$ for annotating a tweet with an arbitrary number of entities and a higher score $5x$ for correctly annotating the pre-computed number of entities (based on the gold standard tags). These figures were drawn from a series of observations and early pilots, which also ruled out the use of negative scores and qualifying questions, as both led to a very sharp rate of exits. As noted earlier, workers had access to a $k$-neighbours leaderboard, which lists contenders immediately above or below. 

\subsubsection{Requester parameters} Wordsmith affords the requester a number of configurable parameters including:

\begin{itemize}
\item \textbf{Number of workers $W$:} count of those who could connect to the platform. All the workers had to be connected before the task starts. 
\item \textbf{Leaderboard}: requesters can choose between listing the $top-k$ workers; or displaying the $k$ contenders above and below the worker. 
\item \textbf{Number of tasks $P$}: this represents the total number of tasks which are to be processed, each consisting on $n$ objects to be annotated.
\item \textbf{Window size $w$}: this is number of tasks which is sent out to a number of workers per time slice (see below).
\item \textbf{Task unit time \begin{math}\mu\end{math}}: this is the delay time, for which a list of tasks remains available for annotation to a worker before the next set of tasks arrive. 
\item \textbf{Task arrival rate \begin{math}\lambda\end{math}}: this variable measures the number of tasks, which are channelled to the platform per unit time. 
\item \textbf{Task intensity $L$} (or ingress load): this is computed as \begin{math}L = \lambda/\mu\end{math}, based on the task arrival rate \begin{math}\lambda\end{math} and the agent processing rate \begin{math}\mu\end{math}.
\item \textbf{Task unit size}: represents the number of tasks that a worker actually sees on screen at any given time, which is a fixed percentage of the window size $w$. 
\item \textbf{Total task time $T$}: computes how long the contest would take as a function of the number of tweets and the window size: \begin{math}T = P\mu/w\end{math}.
\item \textbf{Reward spread $s$}: the reward spread denotes the number of workers to be paid in the current contest.
\end{itemize}

\subsubsection{Warping time} Warping time is a strategy, in which a workers task slice in a real-time assignment is deliberately slowed down to afford for maximal worker cognition in undertaking the required task. For example, \cite{lasecki2013warping} used time warping to slow down audio playback so crowd workers could effectively transcribe a given portion of speech. This was recursively done for each worker, after which the individual results were successively merged to create a single result. Following our stream parallelism approach to dividing up the incoming microposts, a window of $200$ tweets was presented to $20$ workers. 

We adopted the approach by \cite{lasecki2013warping} denoting: (i) an in-period \begin{math}P_i\end{math} where the annotation stream for a worker group comes in; (ii) a speed reduction rate \textit{rr}; and (iii) the compensating out-period \begin{math}P_o\end{math} where the worker group $WG_i$ rejoins the live real-time stream. We used a speed reduction rate \textit{rr} of $10$ i.e., the $20$ workers experienced a streaming rate of \begin{math}1/rr = 0.10\end{math} (i.e. one tenth speed, $10$ seconds rather than $1$ second) while annotating.  During these $10$ seconds, the workers were presented with static list of tweets, from which they could select individual entities to annotate. The stream of tweets in the buffered out period was then emitted at a speed of:
\begin{equation}
\frac{N - 1}{N - rr}
\end{equation}

In our experiments we made certain assumptions, which would be handled differently on a live data feed. For example, all members of a worker group ($20$ workers in our case) were presented with the annotation tasks at the same input period \begin{math}P_o\end{math}, which was a function of our streaming approach. We re-purposed existing datasets from literature into a streaming API. In actual practice, each worker would have a unique input period \begin{math}P_o\end{math} similar to \cite{lasecki2013warping}.

\subsubsection{Task allocation} There are several approaches to distribute the incoming stream of tasks to the available workers in parallel. In a \textit{round robin} assignment, each bin would be sequentially assigned from the window of current annotation tasks to the next available worker, e.g., in our experiments, we assigned each worker $10$ tasks from $1$ out of $20$ bins derived from the streaming window of $200$ tasks.

\section{Experiments}
In this section, we describe how we carry out empirical contest-based crowdsourcing experiments using Wordsmith.

\label{sec:experiments}
\subsection{Datasets}

The dataset consisted of $7,600$ aggregated tweets from four corpora from the literature. These datasets were from different time frames and all come along with gold standards, which we used to perform quality checks and compute contest scores. 


\begin{itemize}
\item \textbf{\textit{The Ritter corpus}} by \cite{ritter2011named} which consists of $2,400$ tweets. The tweets were randomly sampled, however the sampling method and original dataset size are unknown. It is estimated that the tweets were harvested around September $2010$.
\item \textbf{\textit{The Finin corpus}} by \cite{finin2010annotating} consists of $441$ tweets which was the gold standard for a crowdsourcing annotation exercise. It is not stated how the corpus was created, however our investigation puts the corpus between August to September $2008$. 
\item \textbf{\textit{The MSM 2013 corpus}}, the Making Sense of Microposts $2013$ Concept Extraction Challenge dataset by \cite{basave2013making}, which includes training, test, and gold data; for our experiments we used the gold subset comprising $1450$ tweets.
\item \textbf{\textit{The Wordsmith corpus}}, reported in one of our previous works \cite{feyisetan2015towards}. From the a corpus of six billion tweets, we sampled out $3,309$ English ones using \textit{reservoir sampling} - a family of randomized algorithms for sampling $k$ items from a list $S$ of $n$ items.
\end{itemize}

\subsection{Experimental setup}

\subsubsection{Overview}
Our experiments were designed for time-bound microtask crowdsourcing scenarios, which require speedy human-cognitive intervention. The workers were presented with a time-delayed view of streaming tweets. Each experiment competition ran for about six minutes, during which, workers labelled as many tweets as possible, identifying instances of people, locations and organisations. The objective of each worker was to qualify for a set of top reward spots, which guaranteed a payoff. Unlike traditional paid microtasks, only a few of the workers got paid even if they produced correct annotations. This added to the urgency and the competitive drive that sped up the task to near real-time completion rates. We considered several questions, chief of which was: how does the reward spread affect the total task output of all the contest participants. Other questions naturally follow, such as: the relationship between reward spread and annotation time, annotation quality and worker exit. These formed the basis of our experimental conditions.

\subsubsection{Timely worker recruitment strategies}
\label{sec:timely_recruit}
In order to achieve timely worker recruitment, we adopted a combination of several strategies:

\begin{itemize}
\item We posted our tasks repeatedly in order to maintain visibility within the recent tasks view of workers.
\item We created multiple tasks that pointed to our Wordsmith platform, but ensured that workers could connect only once by keeping track of the connection IP address.
\item We  attempted to recruit, on the average, $10$ times the number of workers than we required (around $1,000$ workers). 
\item We posted the tasks in bits as a work around for the scheduling mechanism which CrowdFlower uses in displaying unfinished tasks to new workers.
\item We introduced an acoustic signal to notify workers once the requisite number of contestants had connected to the system.
\end{itemize}

Workers could see in real-time how many more contestants were required to connect before the task started. We also ensured that impatient workers were reconnected whenever they refreshed their screens, however, once the required number of workers were connected, no further connection was allowed.

\subsubsection{Requester parameters} 

In each experimental condition, the number of workers was $100$, viewing a $k=3$ neighbours leaderboard; the number of tasks (in our case Twitter posts) desired to be processed was $7,600$; the window size was $200$; the task unit time was $10$ seconds, the task unit size was $10$. This means that a worker saw $10$ tweets out of the current stream of $200$ tweets for a period of $10$ seconds during which the worker was to label as many as possible before the next set of tasks arrived. The task arrival rate was $200/10$, which means $20$ tweets per second, and the total contest time was $(7,600 * 10)/200$ = $380$ seconds ($6$ minutes $20$ seconds). 


Each contest had a payoff of $\$0.10$ for each prize payment, replicating \cite{feyisetan2015towards,feyisetanextended}. Our model rewarded each winner with the same payment, as opposed to other variants which take into account the ranking of the participants or implement some other form of reward sharing. A reward spread of $1$ stands for a winner-takes-all condition, in which only the top worker gets paid and with a reward spread of $10$, the top ten workers are paid. The parameters have been explained earlier in Sec. \ref{sec:contestdesign}.


\subsection{Experimental design}
\subsubsection{Overview}
We set up our experiment using the between-subjects design. Each worker could only be assigned to one of the experiment conditions. This was enforced using the worker's connection IP address as stated earlier in Sec. \ref{sec:timely_recruit}. Each user was randomly assigned to one of the experiment conditions and they were not able to pre-select which condition they ended in. The multiple tasks posted on CrowdFlower and referenced in Sec. \ref{sec:timely_recruit} served as an identical entry point for each worker. From there, they are redirected to a centralized server that randomly assigns the workers to a certain treatment.

Our experiments had one independent variable: the reward spread, and one dependent variable: the task output. As we described in our theoretical modelling in Sec. \ref{sec:ctmc}, our hypothesis is that the reward spread can `explain' the changes in the contest output (referenced as the total annotations carried out by the workers). Therefore, the variable that a requester seeks to optimize is the total output, and they seek to control this by varying the reward spread.

\begin{itemize}
\item \textbf{Independent variable}: reward spread

\item \textbf{Dependent variable}: task output

\end{itemize}

\subsubsection{Conditions}
The experiments in this paper are compared against a control condition from an identical task reported in \cite{feyisetan2015towards,feyisetanextended}. In that study, workers were also required to annotate named entities in tweets, however, they were not under any time constraint to carry out the task, and every worker that successfully completed the task got paid. In this paper however, the experimental conditions or treatments are based on the different reward spreads that crowd workers are randomly assigned to, with the contest setting inducing the time constraint.

\begin{itemize}
\item \textbf{Control Condition} Pay all workers

\item \textbf{Experimental Condition $1$} Pay top worker

\item \textbf{Experimental Condition $2$} Pay top $5$ workers

\item \textbf{Experimental Condition $3$} Pay top $10$ workers

\end{itemize}

We evaluated task performance in terms of task output (i.e., volume of work done). To assess the task output, we measured the total and average number of texts annotated by the annotators.

\section{Results}
\label{sec:results}
In this section, we discuss the results of our experiment runs, specifically, how the task output from contests yields a significant increase when compared to the baseline control condition. We then discuss the difference in annotation times across the control and experimental conditions, which is a direct consequence of the contest setting. 

Then, we touch on the speed vs. quality trade-off that we incur as an aftereffect of the gain in task output. Finally, we revisit the subject of exits, which comes as a result of potential loss of utility on the part of the crowd workers. We tie this to the reward spread to explain how a change in reward spread can potentially stave off task abandonment.

\begin{figure*}[h]
    \centering
    \begin{subfigure}[t]{0.49\textwidth}
\centering
\includegraphics[width=0.81\columnwidth]{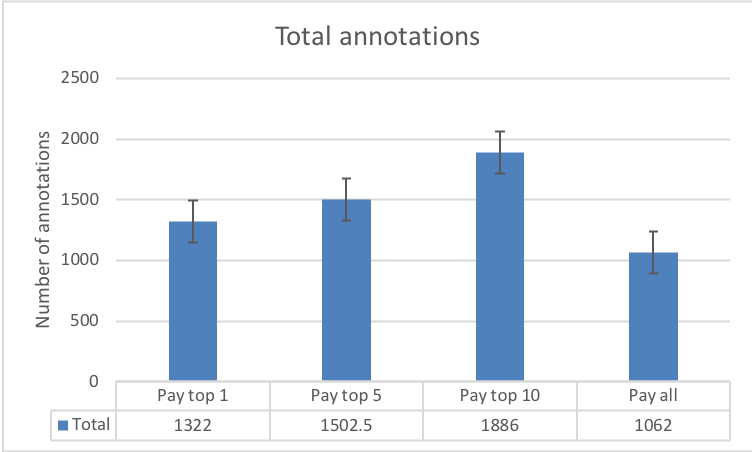}
\caption{Total annotations by all workers}
\label{fig:total_annot}
    \end{subfigure}%
    ~ 
    \begin{subfigure}[t]{0.49\textwidth}
        \centering
    \includegraphics[width=0.81\columnwidth]{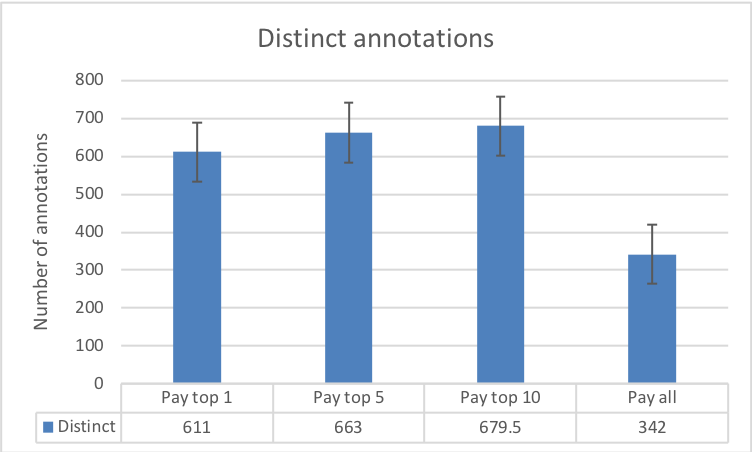}
\caption{Distinct annotations by all workers}
\label{fig:distinct_annot}
    \end{subfigure}
    \\
    \begin{subfigure}[t]{0.49\textwidth}
        \centering
    \includegraphics[width=0.81\columnwidth]{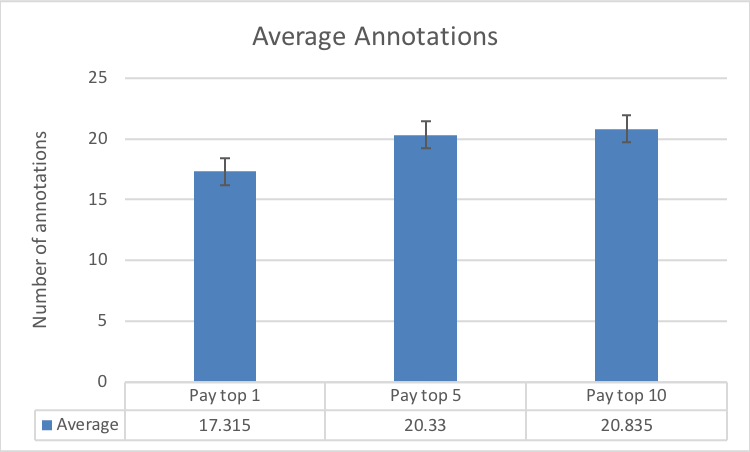}
\caption{Average annotations per worker}
\label{fig:average_annot}
    \end{subfigure}
    ~ 
    \begin{subfigure}[t]{0.49\textwidth}
\centering
\includegraphics[width=0.81\columnwidth]{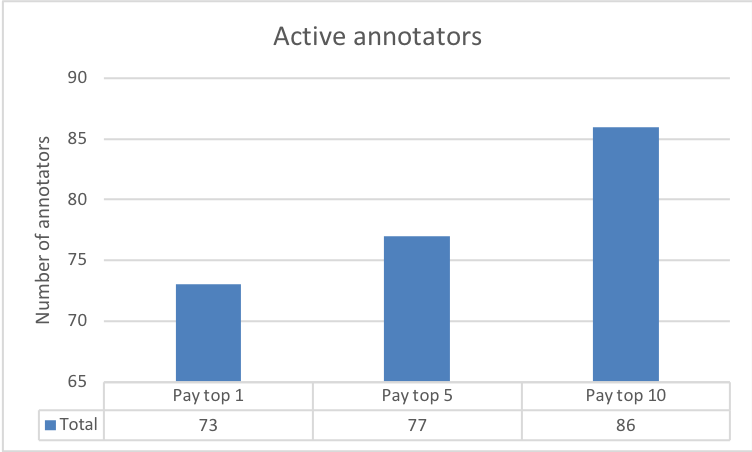}
\caption{Number of active annotators}
\label{fig:active_annot}
    \end{subfigure}%
    \\
        \begin{subfigure}[t]{0.49\textwidth}
\centering
\includegraphics[width=0.81\columnwidth]{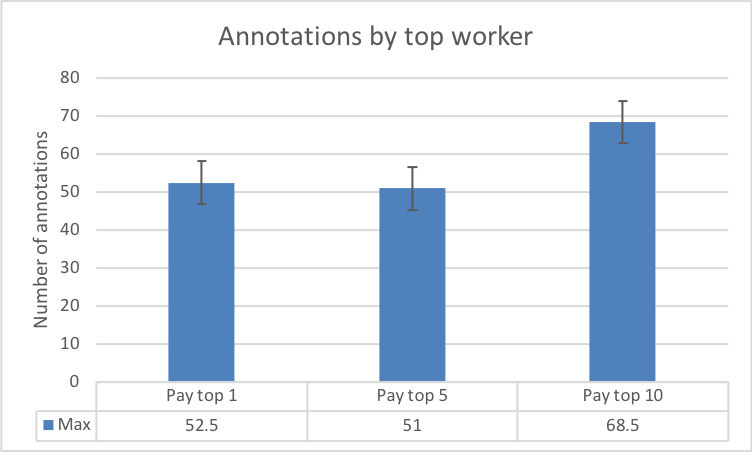}
\caption{Total annotations by top worker}
\label{fig:top_annot}
    \end{subfigure}%
    ~ 
    \begin{subfigure}[t]{0.49\textwidth}
        \centering
    \includegraphics[width=0.81\columnwidth]{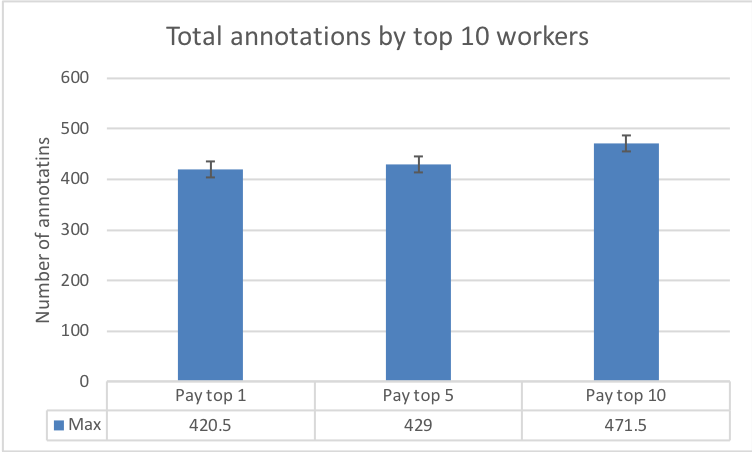}
\caption{Total annotations by top $10$ workers}
\label{fig:top10_annot}
    \end{subfigure}
    
    \caption{Task output}
    \label{fig:task_output}
\end{figure*}

\subsection{Task output}
The results in Fig. \ref{fig:task_output} illustrate how a change in reward spread leads to a change in the number of annotations submitted by workers. The central aim of the requester is to elicit as many annotations from the workers as possible (within the time constraint). Therefore, Fig. \ref{fig:total_annot} and Fig. \ref{fig:distinct_annot} paint a picture of how an increase in the reward spread realises this objective. Both figures compare the total and distinct number of annotations in each of our experimental conditions, against the annotations from the baseline control condition.

From Fig. \ref{fig:total_annot}, the null hypothesis would lead us to believe that the population mean of worker annotations (given by the control condition) is a total of $1,062$ labels in the given time window. The alternative hypothesis from our experimental conditions suggest otherwise: that setting the task as a contest leads to a different mean, which varies by the reward spread. The resulting hypothesis that contests leads to more annotations is statistically significant. A corollary can be drawn from the distinct annotations in Fig. \ref{fig:distinct_annot}.

We then carried out a one-way ANOVA test on the total annotations from the different experimental conditions to see if there was statistical significance as a result of a change in the reward spreads. Across the $3$ conditions: pay top worker ($\mu = 1,322$, $\sigma = 100.40$), pay top 5 workers ($\mu = 1,502$, $\sigma = 132.22$) and pay top 10 workers ($\mu = 1,886$, $\sigma = 134.35$), we obtained results that give evidence to the significance of the results. The F-ratio value is $67.19$, the $p$-value is < $.00001$, and the result is significant at $p$ < $.05$.

Fig. \ref{fig:average_annot} highlights the fairly constant average number of annotations by all the workers. This at first appears to contradict the results we described in Figures \ref{fig:total_annot} and \ref{fig:distinct_annot}. However, Fig. \ref{fig:active_annot} illustrates that the number of active annotators varied across the three experimental conditions. This result also buttresses the significance of the different reward spreads in retaining more annotators. Furthermore, a clearer picture of the number of annotations among workers is evidenced in the the last two Figures \ref{fig:top_annot} and \ref{fig:top10_annot}. The results make plain the effect of varying the reward spread which has stronger impacts on the top workers.

\subsection{Annotation time}
\begin{figure}[ht]
\centering
\includegraphics[width=0.60\columnwidth]{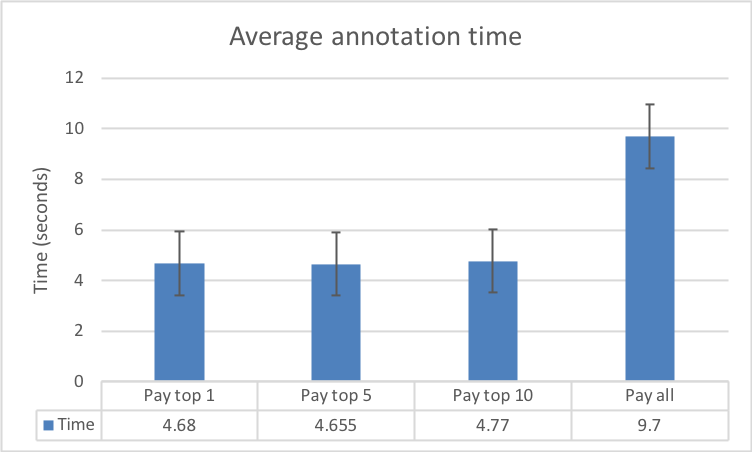}
\caption{Average annotation time per entity}
\label{fig:average_time}
\end{figure}

In Fig. \ref{fig:average_time} we compare the time spent annotating an entity in the different experimental conditions against the time spent by workers in the control condition.

In the control condition, workers were not placed under any specific time constraint, however, inherent timely completion was required to receive their task compensation. Therefore, they still had an incentive to complete the annotations without any delay. In the control condition, workers needed on average $9.70$ seconds to identify and annotate an entity in a tweet. The calculated times in the experimental conditions were as follows: pay top worker ($\mu = 4.68$s, $\sigma=0.08$), pay top $5$ workers ($\mu = 4.65$s, $\sigma=0.32$) and pay top $10$ workers ($\mu = 4.77$s, $\sigma=0.15$).

The results from the contests are significant when compared to the baseline condition which didn't have the explicit time constraint. Contest annotators needed an average of $4.70$s to recognise a type of an entity while regular annotators spent $9.70$s per entity. This is equivalent to an increased annotation factor of almost $2x$ across all contests. Contest annotators spent an average of $5.6$s annotating one tweet (at $1.2$ entities per tweet) and the top contestants averaged $2.69$s per entity, resulting in a gain factor of $3.6x$ against the baseline.

\subsection{Annotation quality}
\begin{figure}[ht]
\centering
\includegraphics[width=0.90\columnwidth]{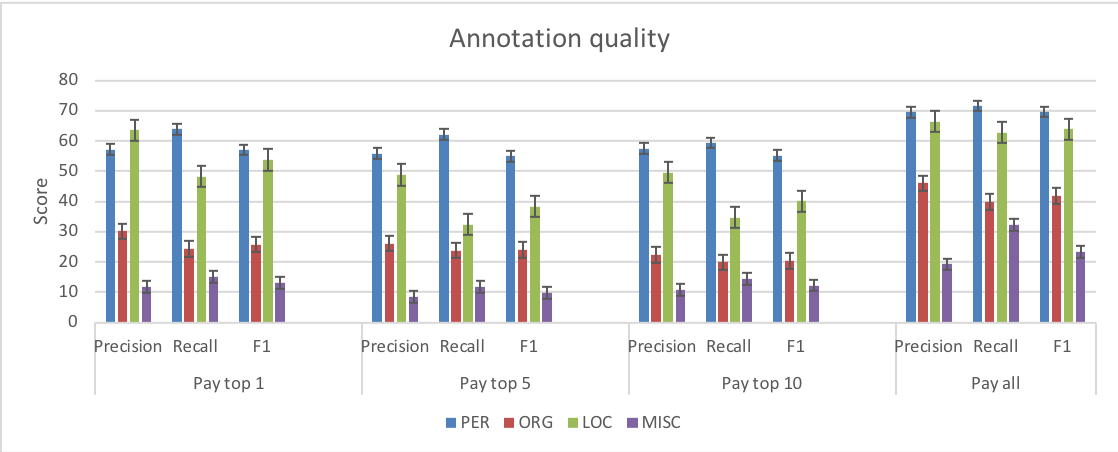}
\caption{Annotation quality: Precision, Recall and F1 scores for annotating Person (PER), Organisation (ORG), Location (LOC) and Miscellaneous (MISC) entities}
\label{fig:annot_quality}
\end{figure}

An increase in annotation speed however led us to incur a hit in annotation quality as reflected in Fig. \ref{fig:annot_quality}. 

The results presented in Fig. \ref{fig:annot_quality} reflect a fine grained analysis of precision, recall and F1 scores on the $4$ entity types (PER, ORG, LOC and MISC) across the experimental and baseline conditions. The results follow the same pattern i.e., the same entity types are equally difficult to identify across all conditions. However, the baseline condition which was free from time constraints resulted in better scores. The scores were computed against the gold standards published with the four datasets used for the experiments.

This result helps inform us as to what tasks contests might be better suited for. A requester seeking to adopt contests needs to be able to tolerate a certain level of false positives and false negatives which would be compensated for by the timeliness of the responses. Verifying the quality of worker submissions (without using inter-annotator agreement scores) however leads to a recursive chicken and egg scenario; for if we were able to automatically compute the quality of the annotations (i.e., by using a gold standard), then what did we need the crowd for? Or better yet, how can we design contests, and assign ranks in a way that creates quality output without falling prey to malicious workers who make quick but incorrect submissions?

In this case, we posit that the task design should be one that satisfies the following two conditions: (a) the worker solution could not have been easily computed automatically; but (b) the worker solution can be computationally verified easily without a gold standard. An example of such a scenario is in the fold.it game \cite{cooper2010predicting} where players fold the structures of selected proteins. Determining a protein's structure is computationally demanding, however, humans are able to intuitively arrive at the best structural composition. Once this has been done, the verification process becomes straightforward making this task a prime candidate for carrying out this form of contest based crowdsourcing without a gold standard.

\subsection{Contest exit behaviour}
\begin{figure}[ht]
\centering
\includegraphics[width=0.80\columnwidth]{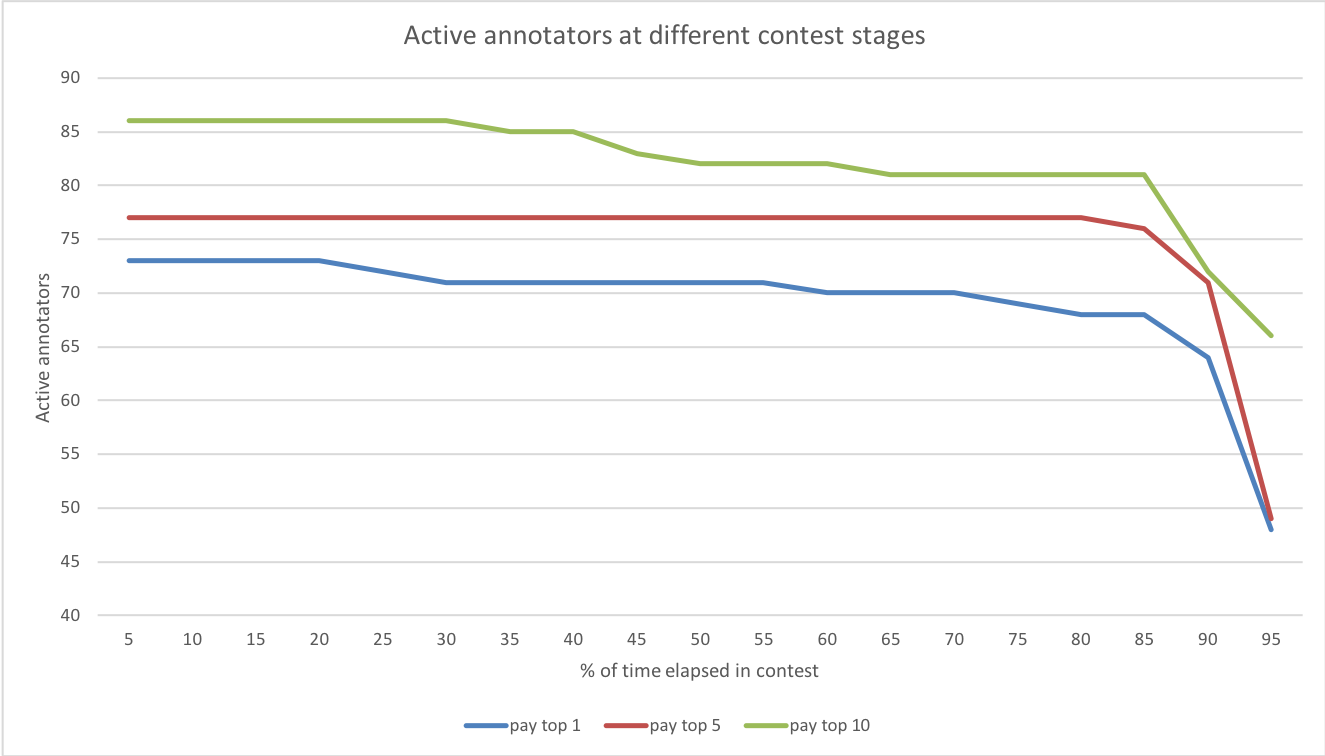}
\caption{Number of active annotators at different stages of the contest. The stages are marked at 5\% time interval steps throughout the contest.}
\label{fig:active_exit}
\end{figure}

Only a subset of contestants received a monetary payoff. The longer workers engaged with a task, the more utility they potentially lost given their ranking relative to the reward spread. Some workers therefore opted to leave the contest prematurely. Figure \ref{fig:active_exit} presents the exit behaviour of workers across the $3$ experimental conditions. The results imply that the beginning of the contests sees fewer exits with most of the contestants choosing to continue for up to $90\%$ of the total time period. The exit rates increase towards the end. This behaviour would be peculiar to microtask contests unlike more longitudinal contests such as presidential elections (studied by \cite{norrander2006attrition}), in which most contenders exit at the beginning of the race leading up to much fewer participants at the final elections. 


\section{Discussion}
\label{sec:discussionandconclusion}
In this section, we highlight lessons learned from the experiments, the challenges of scaling contests to real-time annotations, and the limitations of our approach and results.

\subsection{General considerations}
Crowdsourcing remains a mechanism with tremendous potential to grow economies by engaging a large workforce on demand and at scale. This translates into improved productivity and significant benefits for task requesters and crowd workers alike. Requesters gain task scalability, quick completion and lower price margins for their projects, while workers achieve additional income and learn new skills. In addressing crowdsourcing as a sociotechnical construct, in this paper we focus on one of the crowdsourcing challenges from the list discussed in  \cite{kittur2013future}: real-time crowd work. Tackling this challenge, we developed a model which encourages workers to perform well even in time-sensitive situations by using micropayments and contests.

\subsection{Layering incentives on top of payments} 


Our experiments showed that competitions could work in paid microtask crowdsourcing, both as an alternative to common task models and as a means to complete tasks faster. On average, workers carried out their tasks twice as fast as the baseline workers, while top performers were three times as fast.  When we varied the number of workers who got paid in the experiments from $1$ to $5$ to $10$, we noticed that the average volume of work also went up, with workers perceiving a greater chance to receive part of the available reward. As expected, in all the contest experiments, the gain in speed came at a cost and this was a trade-off for quality. When we compared the quality of submissions in the baseline experiments with that in the contest, we reported a drop in precision of entity identification and total recall of various entity types. In studying the exit patterns of workers from the task, we were keen to understand at what time in the contest do most players quit (potentially due to their inability to potentially finish in the reward spots). Given the intrinsic fun that comes from participation asides from receiving a potential financial payout, many workers (a minimum of $85\%$ per condition) were willing to stay for almost the entire contest time. It was also quite intriguing that $49$ workers rated the contest experience as $3.5$ out of $5$, despite only $9$ of them receiving an actual payout, which makes us believe there must have been some form of satisfaction derived from the competition.  

Our experiments give further insight into the dynamics of crowd behaviour. Workers posted on forums to notify others when a new contest was set up in order to quickly get $100$ workers into the task. This was consistent to our prior image-labelling experiments that used Wordsmith - in \cite{feyisetan2015towards} we noted that some workers even promoted the game to their friends. During the study Wordsmith was used to label images by users who were not recruited via CrowdFlower who found out about the platform via social channels.

\subsection{The role of financial rewards} 



With the contests, we investigated the interplay between the reward spread
and crowd behaviour. Allowing more workers to be eligible for payment improved the overall performance. With more winning spots came an increase in the total and average task output by all participants. Having a higher reward spread also ensured that workers stayed in the context and did not drop out early, thus leading to more tasks completed. This further meant that more effort was required to achieve one of the top spots in the ranking, which would receive a payment.

However, indefinitely increasing the reward spread would defeat the purpose of adopting a contest model, converging towards a traditional microtask system. Our results show that the linear result growth begins to break down as expected at some point. It is important to note that the motivation of crowd workers covers a wide spectrum of intrinsic and extrinsic factors, hence, having a wider reward spread (and a higher task threshold as seen in $C6$) led to a plummet in task output (over repeated experiment runs). An investigation into the discussion forums indicated that this experiment was probably less challenging -- as stated by one worker, "\textit{... to get into the top 10 is not too difficult ...}" - and hence might not have been as attractive to top performers. Understanding where to draw the line would be the subject of further empirical studies paired up with theoretical analysis.

\subsection{Scaling towards real-time annotations} 
Microtask crowdsourcing has often been praised for its ability to produce results quickly and accurately. Yet, an increasing number of applications come in with much harder time constraints, which push the boundaries of the traditional microtask model to deliver in seconds or less \cite{bernstein2011crowds}. Examples of such real-time crowdsourcing applications include machine learning for image recognition \cite{vonAhn:2004:LIC:985692.985733} and text-to-speech conversion \cite{lasecki2013warping}; accessibility design \cite{bigham2010vizwiz,lasecki2012real,lasecki2013warping}; and disaster management \cite{gao2011harnessing}. 

The study lends support to our hypothesis that the microtask contest model leads to timely task completion without the associated overhead costs. Our results show an increase in speed by an average factor of $2x$, and up to $3x$ more among the top performers. These metrics could be used to inform decisions on the number of workers that would be needed to annotate the entire dataset of $7,600$ tweets in real-time - taking into consideration the total number of non-unique annotations and worker exit e.g., in experiment $C5$, workers annotated $1981$ tweets, equivalent to $384$ workers required to make one pass at the entire stream. Our experiments were carried out on a stream of $20$ tweets per second; the live Twitter stream is currently estimated at $6,000$ tweets per second or $2,400$ English tweets per second ($40\%$ of the full stream) \cite{feyisetan2004semantics}. Annotating $10\%$ of the live stream could be potentially carried out by designing a contest for ~$4,600$ workers (i.e., $384$ workers * (($10\%$ of $2,400$) / $20$ tweets per second)).

\subsection{A diverse crowd}

Across all experimental conditions, we have observed a range of responses to the incentives we layered on top of micropayments, which varied depending on a range of factors, including the amount of work already completed and the likelihood to be rewarded. During the contests, top contributors showed a different behaviour in the conditions with 
larger reward spreads. These findings suggest the need for additional experiments, which would investigate the effects of the different incentives on different groups of workers - for example, those who contributed a lot in terms of output, but also those who over achieve in terms of quality. This would potentially allow us to predict exit behaviour more accurately, and offer finer grained incentives - for instance, one could imagine different bonus levels based on how many objects a worker has annotated so far.

\subsection{Application to active learning}
Active learning is a machine learning paradigm that seeks to minimise the time and crowdsourced labels required to train a machine learning model \cite{settles2012active}. Since the objective of active learning is to scale down the the amount of labels required, it is of necessity that the labels be of high quality. However, in large scale real-time active learning and crowdsourcing scenarios \cite{vijayanarasimhan2014large,yan2011active}, a single `oracle' providing labels will not suffice. Just as active learning algorithms refine the model and suggest the few training examples required, crowdsourcing contests refine the annotator pool to provide the `expert crowd' required to improve the model's decision boundary. Applying the contest model in an active learning setting suggests that contest incentives do not always have to optimise for speed (unlike in the real-time scenario). This opens up avenues for future research into incentives in high-value judgement tasks that are aligned more towards annotator skill and less towards completion speed.

\subsection{Payments and ethical considerations} Crowd worker motivation remains a constant research area in understanding why people partake in microtasks. This is essential in order to design systems that are fair and rewarding to the workers. A requester would always seek to minimise cost, however, a worker's complete range of motivations is not yet fully understood. Our investigations revealed that the task model was relatively well received: in one of the baseline experiments, $87$ workers rated the payment of $\$0.10$ as $4$ out of $5$, while in one of our contests, $49$ workers rated it as $3.5$ on the same scale despite only $9$ of them receiving an actual payout. (The numbers $87$ and $49$ represent all the workers who gave feedback on the task). Furthermore, workers seemed to be eager to return - as stated in a crowd discussion forum, one of the participants posted "\textit{Hit the top $10$ today. I will hunt this problem again}", while another one felt let down when they couldn't be among the $100$ contestants: "\textit{Yesterday I came across this, but [they] recruited 100 people, [I] was not allowed to play}".

On task payment, we favoured a higher than average payment of $\$0.10$ as against the annotation averages reported by \cite{difallah2015dynamics}, however, it would be interesting to see the effects of increased payments on the results. In a related set of experiments we noted that raising the reward to as much as $\$0.25$ rather created an anchoring effect than a trend in results. We would like to investigate this further and also analyse the effect of task payment ordering. A greater understanding of worker intrinsic motivations would help in the design of better payment schemes, wrapped around task models that workers find inherently engaging and rewarding; and lend insights into the ongoing debate on ethical and fair crowdsourcing \cite{irani2013turkopticon}.

\subsection{Limitations} 
\label{sec:limitations}
While the results of our experiments were very promising, we are aware of several limitations of our work. The experiments are run on text annotation tasks. It would be interesting to test the models on the other five types of tasks from \cite{gadiraju2014taxonomy}. While our experiments and theories could be easily adapted to similar task types, in particular to output-agreement ones \cite{vonAhn:2004:LIC:985692.985733}, they are less suitable for settings in which a diversity of answers is sought. As a result, there is much latitude for future work across the different incentive schemes and the concept of furtherance incentives as a whole. The tasks we tested do not require a high cognitive load, time investment or creativity. Other types of tasks might be less suitable for a casual game model like the one in Wordsmith. Moreover, the results will probably vary when the tasks are carried out by a community who is intrinsically motivated (for example, to advance science as in citizen science projects, or to help, as in disaster relief). Those motivations would probably overshadow the simpler mechanisms tested here. In addition, such intrinsic motivation settings have been shown to be more prone to overjustification effects, and any attempt to use incentives to reward contributions might have negative effects. 

Just like most empirical research in paid microtask crowdsourcing, none of the experiments considered temporal effects. While contests may produce positive effects short-term, it would be interesting to understand whether workers would continue to engage more with no additional incentives as tasks run over longer periods of time. One common concern related to virtual rewards is that they are not always aligned with the needs and values of the user and that this mismatch becomes more obvious with time \cite{deterding2011game}. In our experiments we have not carried out an in-depth study of the types of workers who used Wordsmith or try to understand what drives them. While the experiments were successful, we anticipate that we would need to have a much more nuanced approach to contest-based incentives when workers engage with us over longer periods of time.

Our contest experiments allowed us to get a sense of the number of contributors required to annotate a substantial share of the data feed in near-real-time. However, our findings might not transfer to scenarios with contests that stretch over days or longer.  In the same time, in a real-world scenario, most human labelling projects that operate with large amounts of data would also include an algorithmic component, trained by labels created by the crowd.

Finally, our work assumes that there is a gold standard to verify crowd answers. This is particularly important in time-sensitive scenarios.

\section{Conclusions and future work}
Our experiments have explored the effects of layered incentive mechanisms on a base financial payment model on crowd performance and engagement. 
The results suggest that we can build better crowdsourcing systems for task requesters and workers alike; and that money need not be the only currency for transacting on paid microtask platforms

In line with the research on timely task completion, we intend to carry out new real-time crowdsourcing experiments which feature automatically verifiable task output which is currently a limitation of our current approach. We would also carry out cross studies between social incentives-based microtasks and contest-based tasks in a bid to understand how the former could be used to facilitate the latter. Our approach to use competitions to deliver timely annotations naturally engenders worker exit given the varying utilities gained by individual workers during the contest. By doing so, it offers us a way to gain insight into how the different parameters we use to our probabilistic exit predictor impact behaviour.

Overall, our results reinforced the role of motivation and incentives beyond cash payments as the primary means to reward participation on some of the most popular crowdsourcing platforms of our times. Our experiments give crowdsourcing researchers and platform designers new empirical evidence into crowdsourcing as a fully fledged sociotechnical construct with challenges requiring as much a social touch as the technical. Worker motivations cover a wide spectrum, which extends beyond the simplistic spectrum of `fun and money'. Understanding how to keep the crowd engaged is essential not just to yield superior returns for requesters, but also to create a more humane experience, as crowdsourcing continues to take its place as the model for the future of work.

\begin{acks}
The work has received funding from the European Union's Horizon 2020 research and innovation programme under grant agreement no 732194 (QROWD), and from the UK EPSRC under grant EP/J017728/2 (SOCIAM: The Theory and Practice of Social Machines). We would also like to acknowledge Karen Hovsepian for all the helpful conversations.
\end{acks}

\bibliographystyle{ACM-Reference-Format}
\bibliography{references}


\begin{thebibliography}{00}


\ifx \showCODEN    \undefined \def \showCODEN     #1{\unskip}     \fi
\ifx \showDOI      \undefined \def \showDOI       #1{#1}\fi
\ifx \showISBNx    \undefined \def \showISBNx     #1{\unskip}     \fi
\ifx \showISBNxiii \undefined \def \showISBNxiii  #1{\unskip}     \fi
\ifx \showISSN     \undefined \def \showISSN      #1{\unskip}     \fi
\ifx \showLCCN     \undefined \def \showLCCN      #1{\unskip}     \fi
\ifx \shownote     \undefined \def \shownote      #1{#1}          \fi
\ifx \showarticletitle \undefined \def \showarticletitle #1{#1}   \fi
\ifx \showURL      \undefined \def \showURL       {\relax}        \fi
\providecommand\bibfield[2]{#2}
\providecommand\bibinfo[2]{#2}
\providecommand\natexlab[1]{#1}
\providecommand\showeprint[2][]{arXiv:#2}

\bibitem[\protect\citeauthoryear{Archak}{Archak}{2010}]%
        {archak2010money}
\bibfield{author}{\bibinfo{person}{Nikolay Archak}.}
  \bibinfo{year}{2010}\natexlab{}.
\newblock \showarticletitle{Money, glory and cheap talk: analyzing strategic
  behavior of contestants in simultaneous crowdsourcing contests on TopCoder.
  com}. In \bibinfo{booktitle}{{\em Proceedings of the 19th international
  conference on World wide web}}. ACM, \bibinfo{pages}{21--30}.
\newblock


\bibitem[\protect\citeauthoryear{Bennett and Lanning}{Bennett and
  Lanning}{2007}]%
        {bennett2007netflix}
\bibfield{author}{\bibinfo{person}{James Bennett} {and} \bibinfo{person}{Stan
  Lanning}.} \bibinfo{year}{2007}\natexlab{}.
\newblock \showarticletitle{The Netflix Prize}. In \bibinfo{booktitle}{{\em
  Proceedings of KDD Cup and Workshop}}, Vol.~\bibinfo{volume}{2007}.
  \bibinfo{pages}{35}.
\newblock


\bibitem[\protect\citeauthoryear{Bernstein, Brandt, Miller, and
  Karger}{Bernstein et~al\mbox{.}}{2011}]%
        {bernstein2011crowds}
\bibfield{author}{\bibinfo{person}{Michael~S Bernstein}, \bibinfo{person}{Joel
  Brandt}, \bibinfo{person}{Robert~C Miller}, {and} \bibinfo{person}{David~R
  Karger}.} \bibinfo{year}{2011}\natexlab{}.
\newblock \showarticletitle{Crowds in two seconds: Enabling real-time
  crowd-powered interfaces}. In \bibinfo{booktitle}{{\em Proceedings of the
  24th annual ACM Symposium on User Interface Software and Technology}}. ACM,
  \bibinfo{pages}{33--42}.
\newblock


\bibitem[\protect\citeauthoryear{Bernstein, Karger, Miller, and
  Brandt}{Bernstein et~al\mbox{.}}{2012}]%
        {bernstein2012analytic}
\bibfield{author}{\bibinfo{person}{Michael~S Bernstein},
  \bibinfo{person}{David~R Karger}, \bibinfo{person}{Robert~C Miller}, {and}
  \bibinfo{person}{Joel Brandt}.} \bibinfo{year}{2012}\natexlab{}.
\newblock \showarticletitle{Analytic methods for optimizing realtime
  crowdsourcing}.
\newblock \bibinfo{journal}{{\em Collective Intelligence\/}}
  (\bibinfo{year}{2012}).
\newblock


\bibitem[\protect\citeauthoryear{Bernstein, Little, Miller, Hartmann, Ackerman,
  Karger, Crowell, and Panovich}{Bernstein et~al\mbox{.}}{2015}]%
        {bernstein2015soylent}
\bibfield{author}{\bibinfo{person}{Michael~S Bernstein}, \bibinfo{person}{Greg
  Little}, \bibinfo{person}{Robert~C Miller}, \bibinfo{person}{Bj{\"o}rn
  Hartmann}, \bibinfo{person}{Mark~S Ackerman}, \bibinfo{person}{David~R
  Karger}, \bibinfo{person}{David Crowell}, {and} \bibinfo{person}{Katrina
  Panovich}.} \bibinfo{year}{2015}\natexlab{}.
\newblock \showarticletitle{Soylent: a word processor with a crowd inside}.
\newblock \bibinfo{journal}{{\it Commun. ACM}} \bibinfo{volume}{58},
  \bibinfo{number}{8} (\bibinfo{year}{2015}), \bibinfo{pages}{85--94}.
\newblock


\bibitem[\protect\citeauthoryear{Bigham, Jayant, Ji, Little, Miller, Miller,
  Miller, Tatarowicz, White, White, et~al\mbox{.}}{Bigham
  et~al\mbox{.}}{2010}]%
        {bigham2010vizwiz}
\bibfield{author}{\bibinfo{person}{Jeffrey~P Bigham},
  \bibinfo{person}{Chandrika Jayant}, \bibinfo{person}{Hanjie Ji},
  \bibinfo{person}{Greg Little}, \bibinfo{person}{Andrew Miller},
  \bibinfo{person}{Robert~C Miller}, \bibinfo{person}{Robin Miller},
  \bibinfo{person}{Aubrey Tatarowicz}, \bibinfo{person}{Brandyn White},
  \bibinfo{person}{Samual White}, {et~al\mbox{.}}}
  \bibinfo{year}{2010}\natexlab{}.
\newblock \showarticletitle{VizWiz: nearly real-time answers to visual
  questions}. In \bibinfo{booktitle}{{\em Proceedings of the 23nd annual ACM
  Symposium on User Interface Software and Technology}}. ACM,
  \bibinfo{pages}{333--342}.
\newblock


\bibitem[\protect\citeauthoryear{Boudreau, Lacetera, and Lakhani}{Boudreau
  et~al\mbox{.}}{2011}]%
        {boudreau2011incentives}
\bibfield{author}{\bibinfo{person}{Kevin~J Boudreau}, \bibinfo{person}{Nicola
  Lacetera}, {and} \bibinfo{person}{Karim~R Lakhani}.}
  \bibinfo{year}{2011}\natexlab{}.
\newblock \showarticletitle{Incentives and problem uncertainty in innovation
  contests: An empirical analysis}.
\newblock \bibinfo{journal}{{\em Management science\/}} \bibinfo{volume}{57},
  \bibinfo{number}{5} (\bibinfo{year}{2011}), \bibinfo{pages}{843--863}.
\newblock


\bibitem[\protect\citeauthoryear{Burkett}{Burkett}{2013}]%
        {randyburkett2013}
\bibfield{author}{\bibinfo{person}{Randy Burkett}.}
  \bibinfo{year}{2013}\natexlab{}.
\newblock \bibinfo{title}{An Alternative Framework for Agent Recruitment: From
  MICE to RASCLS}.
\newblock   (\bibinfo{year}{2013}).
\newblock
\showURL{%
\url{https://calhoun.nps.edu/handle/10945/43831}}


\bibitem[\protect\citeauthoryear{Cano~Basave, Varga, Rowe, Stankovic, and
  Dadzie}{Cano~Basave et~al\mbox{.}}{2013}]%
        {basave2013making}
\bibfield{author}{\bibinfo{person}{Amparo~Elizabeth Cano~Basave},
  \bibinfo{person}{Andrea Varga}, \bibinfo{person}{Matthew Rowe},
  \bibinfo{person}{Milan Stankovic}, {and} \bibinfo{person}{Aba-Sah Dadzie}.}
  \bibinfo{year}{2013}\natexlab{}.
\newblock \showarticletitle{Making Sense of Microposts (\# MSM2013) Concept
  Extraction Challenge.} \bibinfo{pages}{1--15}.
\newblock


\bibitem[\protect\citeauthoryear{Chawla, Hartline, and Sivan}{Chawla
  et~al\mbox{.}}{2015}]%
        {chawla2015optimal}
\bibfield{author}{\bibinfo{person}{Shuchi Chawla}, \bibinfo{person}{Jason~D
  Hartline}, {and} \bibinfo{person}{Balasubramanian Sivan}.}
  \bibinfo{year}{2015}\natexlab{}.
\newblock \showarticletitle{Optimal crowdsourcing contests}.
\newblock \bibinfo{journal}{{\em Games and Economic Behavior\/}}
  (\bibinfo{year}{2015}).
\newblock


\bibitem[\protect\citeauthoryear{Cooper, Khatib, Treuille, Barbero, Lee,
  Beenen, Leaver-Fay, Baker, Popovi{\'c}, et~al\mbox{.}}{Cooper
  et~al\mbox{.}}{2010}]%
        {cooper2010predicting}
\bibfield{author}{\bibinfo{person}{S. Cooper}, \bibinfo{person}{F. Khatib},
  \bibinfo{person}{A. Treuille}, \bibinfo{person}{J. Barbero},
  \bibinfo{person}{J. Lee}, \bibinfo{person}{M. Beenen}, \bibinfo{person}{A.
  Leaver-Fay}, \bibinfo{person}{D. Baker}, \bibinfo{person}{Z. Popovi{\'c}},
  {et~al\mbox{.}}} \bibinfo{year}{2010}\natexlab{}.
\newblock \showarticletitle{Predicting protein structures with a multiplayer
  online game}.
\newblock \bibinfo{journal}{{\em Nature\/}} \bibinfo{volume}{466},
  \bibinfo{number}{7307} (\bibinfo{year}{2010}), \bibinfo{pages}{756--760}.
\newblock


\bibitem[\protect\citeauthoryear{Dai, Rzeszotarski, Paritosh, and Chi}{Dai
  et~al\mbox{.}}{2015}]%
        {dai2015and}
\bibfield{author}{\bibinfo{person}{Peng Dai}, \bibinfo{person}{Jeffrey~M
  Rzeszotarski}, \bibinfo{person}{Praveen Paritosh}, {and}
  \bibinfo{person}{Ed~H Chi}.} \bibinfo{year}{2015}\natexlab{}.
\newblock \showarticletitle{And now for something completely different:
  Improving crowdsourcing workflows with micro-diversions}. In
  \bibinfo{booktitle}{{\em Proceedings of the 18th ACM Conference on Computer
  Supported Cooperative Work \& Social Computing}}. ACM,
  \bibinfo{pages}{628--638}.
\newblock


\bibitem[\protect\citeauthoryear{Dechenaux, Kovenock, and Sheremeta}{Dechenaux
  et~al\mbox{.}}{2015}]%
        {dechenaux2015survey}
\bibfield{author}{\bibinfo{person}{Emmanuel Dechenaux}, \bibinfo{person}{Dan
  Kovenock}, {and} \bibinfo{person}{Roman~M Sheremeta}.}
  \bibinfo{year}{2015}\natexlab{}.
\newblock \showarticletitle{A survey of experimental research on contests,
  all-pay auctions and tournaments}.
\newblock \bibinfo{journal}{{\em Experimental Economics\/}}
  \bibinfo{volume}{18}, \bibinfo{number}{4} (\bibinfo{year}{2015}),
  \bibinfo{pages}{609--669}.
\newblock


\bibitem[\protect\citeauthoryear{Deci and Ryan}{Deci and Ryan}{2010}]%
        {sdttheory}
\bibfield{author}{\bibinfo{person}{Edward~L. Deci} {and}
  \bibinfo{person}{Richard~M. Ryan}.} \bibinfo{year}{2010}\natexlab{}.
\newblock \bibinfo{booktitle}{{\em Self-Determination}}.
\newblock \bibinfo{publisher}{John Wiley and Sons, Inc.}
\newblock
\showISBNx{9780470479216}


\bibitem[\protect\citeauthoryear{Demartini, Difallah, and
  Cudr{\'e}-Mauroux}{Demartini et~al\mbox{.}}{2013}]%
        {demartini2013large}
\bibfield{author}{\bibinfo{person}{Gianluca Demartini},
  \bibinfo{person}{Djellel~Eddine Difallah}, {and} \bibinfo{person}{Philippe
  Cudr{\'e}-Mauroux}.} \bibinfo{year}{2013}\natexlab{}.
\newblock \showarticletitle{Large-scale linked data integration using
  probabilistic reasoning and crowdsourcing}.
\newblock \bibinfo{journal}{{\em The VLDB Journal\/}} \bibinfo{volume}{22},
  \bibinfo{number}{5} (\bibinfo{year}{2013}), \bibinfo{pages}{665--687}.
\newblock


\bibitem[\protect\citeauthoryear{Deterding, Dixon, Khaled, and Nacke}{Deterding
  et~al\mbox{.}}{2011}]%
        {deterding2011game}
\bibfield{author}{\bibinfo{person}{Sebastian Deterding}, \bibinfo{person}{Dan
  Dixon}, \bibinfo{person}{Rilla Khaled}, {and} \bibinfo{person}{Lennart
  Nacke}.} \bibinfo{year}{2011}\natexlab{}.
\newblock \showarticletitle{From game design elements to gamefulness: defining
  gamification}. In \bibinfo{booktitle}{{\em Proceedings of the 15th
  International Academic MindTrek Conference: Envisioning Future Media
  Environments}}. ACM, \bibinfo{pages}{9--15}.
\newblock


\bibitem[\protect\citeauthoryear{Difallah, Catasta, Demartini, Ipeirotis, and
  Cudr{\'e}-Mauroux}{Difallah et~al\mbox{.}}{2015}]%
        {difallah2015dynamics}
\bibfield{author}{\bibinfo{person}{Djellel~Eddine Difallah},
  \bibinfo{person}{Michele Catasta}, \bibinfo{person}{Gianluca Demartini},
  \bibinfo{person}{Panagiotis~G Ipeirotis}, {and} \bibinfo{person}{Philippe
  Cudr{\'e}-Mauroux}.} \bibinfo{year}{2015}\natexlab{}.
\newblock \showarticletitle{The Dynamics of Micro-Task Crowdsourcing: The Case
  of Amazon MTurk}. In \bibinfo{booktitle}{{\em Proceedings of the 24th
  International Conference on World Wide Web}}. \bibinfo{pages}{238--247}.
\newblock


\bibitem[\protect\citeauthoryear{Feyisetan, Luczak-R{\"o}sch, Simperl, Tinati,
  and Shadbolt}{Feyisetan et~al\mbox{.}}{2015}]%
        {feyisetan2015towards}
\bibfield{author}{\bibinfo{person}{Oluwaseyi Feyisetan},
  \bibinfo{person}{Markus Luczak-R{\"o}sch}, \bibinfo{person}{Elena Simperl},
  \bibinfo{person}{Ramine Tinati}, {and} \bibinfo{person}{Nigel Shadbolt}.}
  \bibinfo{year}{2015}\natexlab{}.
\newblock \showarticletitle{{Towards Hybrid NER: A Study of Content and
  Crowdsourcing-Related Performance Factors}}.
\newblock In \bibinfo{booktitle}{{\em The Semantic Web. Latest Advances and New
  Domains}}. \bibinfo{publisher}{Springer}, \bibinfo{pages}{525--540}.
\newblock


\bibitem[\protect\citeauthoryear{Feyisetan and Simperl}{Feyisetan and
  Simperl}{2016}]%
        {feyisetan2016please}
\bibfield{author}{\bibinfo{person}{Oluwaseyi Feyisetan} {and}
  \bibinfo{person}{Elena Simperl}.} \bibinfo{year}{2016}\natexlab{}.
\newblock \showarticletitle{Please stay vs let’s play: Social pressure
  incentives in paid collaborative crowdsourcing}. In \bibinfo{booktitle}{{\em
  International Conference on Web Engineering}}. Springer,
  \bibinfo{pages}{405--412}.
\newblock


\bibitem[\protect\citeauthoryear{Feyisetan and Simperl}{Feyisetan and
  Simperl}{2017}]%
        {feyisetan2017social}
\bibfield{author}{\bibinfo{person}{Oluwaseyi Feyisetan} {and}
  \bibinfo{person}{Elena Simperl}.} \bibinfo{year}{2017}\natexlab{}.
\newblock \showarticletitle{Social Incentives in Paid Collaborative
  Crowdsourcing}.
\newblock \bibinfo{journal}{{\em ACM Transactions on Intelligent Systems and
  Technology (TIST)\/}} \bibinfo{volume}{8}, \bibinfo{number}{6}
  (\bibinfo{year}{2017}), \bibinfo{pages}{Article no 73}.
\newblock


\bibitem[\protect\citeauthoryear{Feyisetan, Simperl, Luczak-Roesch, Tinati, and
  Shadbolt}{Feyisetan et~al\mbox{.}}{}]%
        {feyisetanextended}
\bibfield{author}{\bibinfo{person}{Oluwaseyi Feyisetan}, \bibinfo{person}{Elena
  Simperl}, \bibinfo{person}{Markus Luczak-Roesch}, \bibinfo{person}{Ramine
  Tinati}, {and} \bibinfo{person}{Nigel Shadbolt}.}
\newblock \showarticletitle{An extended study of content and
  crowdsourcing-related performance factors in named entity annotation}.
\newblock \bibinfo{journal}{{\em Semantic Web\/}} \bibinfo{number}{Preprint}
  (\bibinfo{year}{????}), \bibinfo{pages}{1--25}.
\newblock


\bibitem[\protect\citeauthoryear{Feyisetan, Simperl, Tinati, Luczak-Roesch, and
  Shadbolt}{Feyisetan et~al\mbox{.}}{2014}]%
        {feyisetan2004semantics}
\bibfield{author}{\bibinfo{person}{Oluwaseyi Feyisetan}, \bibinfo{person}{Elena
  Simperl}, \bibinfo{person}{Ramine Tinati}, \bibinfo{person}{Markus
  Luczak-Roesch}, {and} \bibinfo{person}{Nigel Shadbolt}.}
  \bibinfo{year}{2014}\natexlab{}.
\newblock \showarticletitle{Quick-and-clean Extraction of Linked Data Entities
  from Microblogs}. In \bibinfo{booktitle}{{\em Proceedings of the 10th
  International Conference on Semantic Systems}} {\em (\bibinfo{series}{SEM
  '14})}. \bibinfo{publisher}{ACM}, \bibinfo{pages}{5--12}.
\newblock


\bibitem[\protect\citeauthoryear{Feyisetan, Simperl, Van~Kleek, and
  Shadbolt}{Feyisetan et~al\mbox{.}}{2015}]%
        {feyisetan2015improving}
\bibfield{author}{\bibinfo{person}{Oluwaseyi Feyisetan}, \bibinfo{person}{Elena
  Simperl}, \bibinfo{person}{Max Van~Kleek}, {and} \bibinfo{person}{Nigel
  Shadbolt}.} \bibinfo{year}{2015}\natexlab{}.
\newblock \showarticletitle{Improving paid microtasks through gamification and
  adaptive furtherance incentives}. In \bibinfo{booktitle}{{\em Proceedings of
  the 24th International Conference on World Wide Web}}.
  \bibinfo{pages}{333--343}.
\newblock


\bibitem[\protect\citeauthoryear{Finin, Murnane, Karandikar, Keller, Martineau,
  and Dredze}{Finin et~al\mbox{.}}{2010}]%
        {finin2010annotating}
\bibfield{author}{\bibinfo{person}{Tim Finin}, \bibinfo{person}{Will Murnane},
  \bibinfo{person}{Anand Karandikar}, \bibinfo{person}{Nicholas Keller},
  \bibinfo{person}{Justin Martineau}, {and} \bibinfo{person}{Mark Dredze}.}
  \bibinfo{year}{2010}\natexlab{}.
\newblock \showarticletitle{{Annotating named entities in Twitter data with
  crowdsourcing}}. In \bibinfo{booktitle}{{\em Proceedings of the NAACL HLT
  2010 Workshop on Creating Speech and Language Data with Amazon's Mechanical
  Turk}}. Association for Computational Linguistics, \bibinfo{pages}{80--88}.
\newblock


\bibitem[\protect\citeauthoryear{Frei}{Frei}{2009}]%
        {frei2009paid}
\bibfield{author}{\bibinfo{person}{Brent Frei}.}
  \bibinfo{year}{2009}\natexlab{}.
\newblock \showarticletitle{Paid Crowdsourcing}.
\newblock \bibinfo{journal}{{\em Current State \& Progress toward Mainstream
  Business Use, Smartsheet. com Report, Smartsheet. com\/}}
  \bibinfo{volume}{9} (\bibinfo{year}{2009}).
\newblock


\bibitem[\protect\citeauthoryear{Frey and Jegen}{Frey and Jegen}{2001}]%
        {frey2001motivation}
\bibfield{author}{\bibinfo{person}{Bruno~S Frey} {and} \bibinfo{person}{Reto
  Jegen}.} \bibinfo{year}{2001}\natexlab{}.
\newblock \showarticletitle{Motivation crowding theory}.
\newblock \bibinfo{journal}{{\em Journal of economic surveys\/}}
  \bibinfo{volume}{15}, \bibinfo{number}{5} (\bibinfo{year}{2001}),
  \bibinfo{pages}{589--611}.
\newblock


\bibitem[\protect\citeauthoryear{Fudenberg and Tirole}{Fudenberg and
  Tirole}{1986}]%
        {fudenberg1986theory}
\bibfield{author}{\bibinfo{person}{Drew Fudenberg} {and} \bibinfo{person}{Jean
  Tirole}.} \bibinfo{year}{1986}\natexlab{}.
\newblock \showarticletitle{A theory of exit in duopoly}.
\newblock \bibinfo{journal}{{\em Econometrica: Journal of the Econometric
  Society\/}} (\bibinfo{year}{1986}), \bibinfo{pages}{943--960}.
\newblock


\bibitem[\protect\citeauthoryear{Gadiraju, Kawase, and Dietze}{Gadiraju
  et~al\mbox{.}}{2014}]%
        {gadiraju2014taxonomy}
\bibfield{author}{\bibinfo{person}{Ujwal Gadiraju}, \bibinfo{person}{Ricardo
  Kawase}, {and} \bibinfo{person}{Stefan Dietze}.}
  \bibinfo{year}{2014}\natexlab{}.
\newblock \showarticletitle{A taxonomy of microtasks on the web}. In
  \bibinfo{booktitle}{{\em Proceedings of the 25th ACM Conference on Hypertext
  and Social Media}}. ACM, \bibinfo{pages}{218--223}.
\newblock


\bibitem[\protect\citeauthoryear{Gao, Barbier, and Goolsby}{Gao
  et~al\mbox{.}}{2011}]%
        {gao2011harnessing}
\bibfield{author}{\bibinfo{person}{Huiji Gao}, \bibinfo{person}{Geoffrey
  Barbier}, {and} \bibinfo{person}{Rebecca Goolsby}.}
  \bibinfo{year}{2011}\natexlab{}.
\newblock \showarticletitle{Harnessing the crowdsourcing power of social media
  for disaster relief}.
\newblock \bibinfo{journal}{{\em IEEE Intelligent Systems\/}}
  \bibinfo{number}{3} (\bibinfo{year}{2011}), \bibinfo{pages}{10--14}.
\newblock


\bibitem[\protect\citeauthoryear{Good and Su}{Good and Su}{2011}]%
        {good2011games}
\bibfield{author}{\bibinfo{person}{Benjamin~M Good} {and}
  \bibinfo{person}{Andrew~I Su}.} \bibinfo{year}{2011}\natexlab{}.
\newblock \showarticletitle{Games with a scientific purpose}.
\newblock \bibinfo{journal}{{\em Genome Biology\/}} \bibinfo{volume}{12},
  \bibinfo{number}{12} (\bibinfo{year}{2011}), \bibinfo{pages}{1}.
\newblock


\bibitem[\protect\citeauthoryear{Howe}{Howe}{2006}]%
        {howe2006rise}
\bibfield{author}{\bibinfo{person}{Jeff Howe}.}
  \bibinfo{year}{2006}\natexlab{}.
\newblock \showarticletitle{The rise of crowdsourcing}.
\newblock \bibinfo{journal}{{\em Wired Magazine\/}} \bibinfo{volume}{14},
  \bibinfo{number}{6} (\bibinfo{year}{2006}), \bibinfo{pages}{1--4}.
\newblock


\bibitem[\protect\citeauthoryear{Irani and Silberman}{Irani and
  Silberman}{2013}]%
        {irani2013turkopticon}
\bibfield{author}{\bibinfo{person}{Lilly~C Irani} {and} \bibinfo{person}{M
  Silberman}.} \bibinfo{year}{2013}\natexlab{}.
\newblock \showarticletitle{Turkopticon: Interrupting worker invisibility in
  amazon mechanical turk}. In \bibinfo{booktitle}{{\em Proceedings of the
  SIGCHI Conference on Human Factors in Computing Systems}}. ACM,
  \bibinfo{pages}{611--620}.
\newblock


\bibitem[\protect\citeauthoryear{Kaufmann, Schulze, and Veit}{Kaufmann
  et~al\mbox{.}}{2011}]%
        {kaufmann2011more}
\bibfield{author}{\bibinfo{person}{Nicolas Kaufmann}, \bibinfo{person}{Thimo
  Schulze}, {and} \bibinfo{person}{Daniel Veit}.}
  \bibinfo{year}{2011}\natexlab{}.
\newblock \showarticletitle{More than fun and money. Worker Motivation in
  Crowdsourcing-A Study on Mechanical Turk.}. In \bibinfo{booktitle}{{\em
  AMCIS}}, Vol.~\bibinfo{volume}{11}. \bibinfo{pages}{1--11}.
\newblock


\bibitem[\protect\citeauthoryear{Kittur}{Kittur}{2010}]%
        {kittur2010crowdsourcing}
\bibfield{author}{\bibinfo{person}{Aniket Kittur}.}
  \bibinfo{year}{2010}\natexlab{}.
\newblock \showarticletitle{Crowdsourcing, collaboration and creativity}.
\newblock \bibinfo{journal}{{\em ACM Crossroads\/}} \bibinfo{volume}{17},
  \bibinfo{number}{2} (\bibinfo{year}{2010}), \bibinfo{pages}{22--26}.
\newblock


\bibitem[\protect\citeauthoryear{Kittur, Nickerson, Bernstein, Gerber, Shaw,
  Zimmerman, Lease, and Horton}{Kittur et~al\mbox{.}}{2013}]%
        {kittur2013future}
\bibfield{author}{\bibinfo{person}{Aniket Kittur}, \bibinfo{person}{Jeffrey~V
  Nickerson}, \bibinfo{person}{Michael Bernstein}, \bibinfo{person}{Elizabeth
  Gerber}, \bibinfo{person}{Aaron Shaw}, \bibinfo{person}{John Zimmerman},
  \bibinfo{person}{Matt Lease}, {and} \bibinfo{person}{John Horton}.}
  \bibinfo{year}{2013}\natexlab{}.
\newblock \showarticletitle{The future of crowd work}. In
  \bibinfo{booktitle}{{\em Proceedings of the ACM Conference on Computer
  Supported Cooperative Work}}. ACM, \bibinfo{pages}{1301--1318}.
\newblock


\bibitem[\protect\citeauthoryear{Kittur, Smus, Khamkar, and Kraut}{Kittur
  et~al\mbox{.}}{2011}]%
        {kittur2011crowdforge}
\bibfield{author}{\bibinfo{person}{Aniket Kittur}, \bibinfo{person}{Boris
  Smus}, \bibinfo{person}{Susheel Khamkar}, {and} \bibinfo{person}{Robert~E
  Kraut}.} \bibinfo{year}{2011}\natexlab{}.
\newblock \showarticletitle{Crowdforge: Crowdsourcing complex work}. In
  \bibinfo{booktitle}{{\em Proceedings of the 24th annual ACM Symposium on User
  Interface Software and Technology}}. ACM, \bibinfo{pages}{43--52}.
\newblock


\bibitem[\protect\citeauthoryear{Kraut, Resnick, Kiesler, Burke, Chen, Kittur,
  Konstan, Ren, and Riedl}{Kraut et~al\mbox{.}}{2012}]%
        {kraut2012building}
\bibfield{author}{\bibinfo{person}{Robert~E Kraut}, \bibinfo{person}{Paul
  Resnick}, \bibinfo{person}{Sara Kiesler}, \bibinfo{person}{Moira Burke},
  \bibinfo{person}{Yan Chen}, \bibinfo{person}{Niki Kittur},
  \bibinfo{person}{Joseph Konstan}, \bibinfo{person}{Yuqing Ren}, {and}
  \bibinfo{person}{John Riedl}.} \bibinfo{year}{2012}\natexlab{}.
\newblock \bibinfo{booktitle}{{\em Building successful online communities:
  Evidence-based social design}}.
\newblock \bibinfo{publisher}{Mit Press}.
\newblock


\bibitem[\protect\citeauthoryear{Krishna and Morgan}{Krishna and
  Morgan}{1997}]%
        {krishna1997analysis}
\bibfield{author}{\bibinfo{person}{Vijay Krishna} {and} \bibinfo{person}{John
  Morgan}.} \bibinfo{year}{1997}\natexlab{}.
\newblock \showarticletitle{An analysis of the war of attrition and the all-pay
  auction}.
\newblock \bibinfo{journal}{{\em Journal of Economic Theory\/}}
  \bibinfo{volume}{72}, \bibinfo{number}{2} (\bibinfo{year}{1997}),
  \bibinfo{pages}{343--362}.
\newblock


\bibitem[\protect\citeauthoryear{Lasecki, Homan, and Bigham}{Lasecki
  et~al\mbox{.}}{2014}]%
        {lasecki2014architecting}
\bibfield{author}{\bibinfo{person}{Walter~S Lasecki},
  \bibinfo{person}{Christopher Homan}, {and} \bibinfo{person}{Jeffrey~P
  Bigham}.} \bibinfo{year}{2014}\natexlab{}.
\newblock \showarticletitle{Architecting real-time crowd-powered systems}.
\newblock \bibinfo{journal}{{\em Human Computation\/}} \bibinfo{volume}{1},
  \bibinfo{number}{1} (\bibinfo{year}{2014}).
\newblock


\bibitem[\protect\citeauthoryear{Lasecki, Miller, Sadilek, Abumoussa, Borrello,
  Kushalnagar, and Bigham}{Lasecki et~al\mbox{.}}{2012}]%
        {lasecki2012real}
\bibfield{author}{\bibinfo{person}{Walter~S Lasecki},
  \bibinfo{person}{Christopher Miller}, \bibinfo{person}{Adam Sadilek},
  \bibinfo{person}{Andrew Abumoussa}, \bibinfo{person}{Donato Borrello},
  \bibinfo{person}{Raja Kushalnagar}, {and} \bibinfo{person}{Jeffrey Bigham}.}
  \bibinfo{year}{2012}\natexlab{}.
\newblock \showarticletitle{Real-time captioning by groups of non-experts}. In
  \bibinfo{booktitle}{{\em Proceedings of the 25th annual ACM Symposium on User
  Interface Software and Technology}}. ACM, \bibinfo{pages}{23--34}.
\newblock


\bibitem[\protect\citeauthoryear{Lasecki, Miller, and Bigham}{Lasecki
  et~al\mbox{.}}{2013}]%
        {lasecki2013warping}
\bibfield{author}{\bibinfo{person}{Walter~S Lasecki},
  \bibinfo{person}{Christopher~D Miller}, {and} \bibinfo{person}{Jeffrey~P
  Bigham}.} \bibinfo{year}{2013}\natexlab{}.
\newblock \showarticletitle{Warping time for more effective real-time
  crowdsourcing}. In \bibinfo{booktitle}{{\em Proceedings of the SIGCHI
  Conference on Human Factors in Computing Systems}}. ACM,
  \bibinfo{pages}{2033--2036}.
\newblock


\bibitem[\protect\citeauthoryear{Lasecki, Murray, White, Miller, and
  Bigham}{Lasecki et~al\mbox{.}}{2011}]%
        {lasecki2011real}
\bibfield{author}{\bibinfo{person}{Walter~S Lasecki}, \bibinfo{person}{Kyle~I
  Murray}, \bibinfo{person}{Samuel White}, \bibinfo{person}{Robert~C Miller},
  {and} \bibinfo{person}{Jeffrey~P Bigham}.} \bibinfo{year}{2011}\natexlab{}.
\newblock \showarticletitle{Real-time crowd control of existing interfaces}. In
  \bibinfo{booktitle}{{\em Proceedings of the 24th annual ACM Symposium on User
  Interface Software and Technology}}. ACM, \bibinfo{pages}{23--32}.
\newblock


\bibitem[\protect\citeauthoryear{Law, Yin, Joslin~Goh, Terry, and Gajos}{Law
  et~al\mbox{.}}{2016}]%
        {law2016curiosity}
\bibfield{author}{\bibinfo{person}{Edith Law}, \bibinfo{person}{Ming Yin},
  \bibinfo{person}{Kevin~Chen Joslin~Goh}, \bibinfo{person}{Michael Terry},
  {and} \bibinfo{person}{Krzysztof~Z Gajos}.} \bibinfo{year}{2016}\natexlab{}.
\newblock \showarticletitle{Curiosity Killed the Cat, but Makes Crowdwork
  Better}. CHI.
\newblock


\bibitem[\protect\citeauthoryear{Liu, Yang, Adamic, and Chen}{Liu
  et~al\mbox{.}}{2011}]%
        {liu2011crowdsourcing}
\bibfield{author}{\bibinfo{person}{Tracy~Xiao Liu}, \bibinfo{person}{Jiang
  Yang}, \bibinfo{person}{Lada~A Adamic}, {and} \bibinfo{person}{Yan Chen}.}
  \bibinfo{year}{2011}\natexlab{}.
\newblock \showarticletitle{Crowdsourcing with all-pay auctions: A field
  experiment on Taskcn}.
\newblock \bibinfo{journal}{{\em Proceedings of the Association for Information
  Science and Technology\/}} \bibinfo{volume}{48}, \bibinfo{number}{1}
  (\bibinfo{year}{2011}), \bibinfo{pages}{1--4}.
\newblock


\bibitem[\protect\citeauthoryear{Malone, Laubacher, and Dellarocas}{Malone
  et~al\mbox{.}}{2010}]%
        {malone2010collective}
\bibfield{author}{\bibinfo{person}{Thomas~W Malone}, \bibinfo{person}{Robert
  Laubacher}, {and} \bibinfo{person}{Chrysanthos Dellarocas}.}
  \bibinfo{year}{2010}\natexlab{}.
\newblock \showarticletitle{The Collective Intelligence Genome}.
\newblock \bibinfo{journal}{{\em MIT Sloan Management Review\/}}
  \bibinfo{volume}{51}, \bibinfo{number}{3} (\bibinfo{year}{2010}),
  \bibinfo{pages}{21}.
\newblock


\bibitem[\protect\citeauthoryear{Martin, Hanrahan, O'Neill, and Gupta}{Martin
  et~al\mbox{.}}{2014}]%
        {martin2014being}
\bibfield{author}{\bibinfo{person}{David Martin}, \bibinfo{person}{Benjamin~V
  Hanrahan}, \bibinfo{person}{Jacki O'Neill}, {and} \bibinfo{person}{Neha
  Gupta}.} \bibinfo{year}{2014}\natexlab{}.
\newblock \showarticletitle{Being a turker}. In \bibinfo{booktitle}{{\em
  Proceedings of the 17th ACM Conference on Computer Supported Cooperative Work
  \& Social Computing}}. ACM, \bibinfo{pages}{224--235}.
\newblock


\bibitem[\protect\citeauthoryear{Mason and Watts}{Mason and Watts}{2010}]%
        {mason2010financial}
\bibfield{author}{\bibinfo{person}{Winter Mason} {and}
  \bibinfo{person}{Duncan~J Watts}.} \bibinfo{year}{2010}\natexlab{}.
\newblock \showarticletitle{Financial incentives and the performance of
  crowds}.
\newblock \bibinfo{journal}{{\em ACM SigKDD Explorations Newsletter\/}}
  \bibinfo{volume}{11}, \bibinfo{number}{2} (\bibinfo{year}{2010}),
  \bibinfo{pages}{100--108}.
\newblock


\bibitem[\protect\citeauthoryear{Mekler, Br{\"u}hlmann, Opwis, and Tuch}{Mekler
  et~al\mbox{.}}{2013}]%
        {mekler2013disassembling}
\bibfield{author}{\bibinfo{person}{Elisa~D Mekler}, \bibinfo{person}{Florian
  Br{\"u}hlmann}, \bibinfo{person}{Klaus Opwis}, {and}
  \bibinfo{person}{Alexandre~N Tuch}.} \bibinfo{year}{2013}\natexlab{}.
\newblock \showarticletitle{Disassembling gamification: the effects of points
  and meaning on user motivation and performance}. In \bibinfo{booktitle}{{\em
  CHI'13 extended abstracts on human factors in computing systems}}. ACM,
  \bibinfo{pages}{1137--1142}.
\newblock


\bibitem[\protect\citeauthoryear{Michelucci}{Michelucci}{2016}]%
        {michelucci2013handbook}
\bibfield{author}{\bibinfo{person}{Pietro Michelucci}.}
  \bibinfo{year}{2016}\natexlab{}.
\newblock \bibinfo{booktitle}{{\em Handbook of Human Computation}}.
\newblock \bibinfo{publisher}{Springer}.
\newblock


\bibitem[\protect\citeauthoryear{Moldovanu and Sela}{Moldovanu and
  Sela}{2001}]%
        {moldovanu2001optimal}
\bibfield{author}{\bibinfo{person}{Benny Moldovanu} {and} \bibinfo{person}{Aner
  Sela}.} \bibinfo{year}{2001}\natexlab{}.
\newblock \showarticletitle{The optimal allocation of prizes in contests}.
\newblock \bibinfo{journal}{{\em American Economic Review\/}}
  (\bibinfo{year}{2001}), \bibinfo{pages}{542--558}.
\newblock


\bibitem[\protect\citeauthoryear{Moldovanu, Sela, and Shi}{Moldovanu
  et~al\mbox{.}}{2012}]%
        {moldovanu2012carrots}
\bibfield{author}{\bibinfo{person}{Benny Moldovanu}, \bibinfo{person}{Aner
  Sela}, {and} \bibinfo{person}{Xianwen Shi}.} \bibinfo{year}{2012}\natexlab{}.
\newblock \showarticletitle{Carrots and sticks: prizes and punishments in
  contests}.
\newblock \bibinfo{journal}{{\em Economic Inquiry\/}} \bibinfo{volume}{50},
  \bibinfo{number}{2} (\bibinfo{year}{2012}), \bibinfo{pages}{453--462}.
\newblock


\bibitem[\protect\citeauthoryear{Norrander}{Norrander}{2006}]%
        {norrander2006attrition}
\bibfield{author}{\bibinfo{person}{B. Norrander}.}
  \bibinfo{year}{2006}\natexlab{}.
\newblock \showarticletitle{The attrition game: Initial resources, initial
  contests and the exit of candidates during the US presidential primary
  season}.
\newblock \bibinfo{journal}{{\em British Journal of Political Science\/}}
  \bibinfo{volume}{36}, \bibinfo{number}{03} (\bibinfo{year}{2006}),
  \bibinfo{pages}{487--507}.
\newblock


\bibitem[\protect\citeauthoryear{Nov, Arazy, and Anderson}{Nov
  et~al\mbox{.}}{2014}]%
        {nov2014scientists}
\bibfield{author}{\bibinfo{person}{Oded Nov}, \bibinfo{person}{Ofer Arazy},
  {and} \bibinfo{person}{David Anderson}.} \bibinfo{year}{2014}\natexlab{}.
\newblock \showarticletitle{Scientists@ Home: what drives the quantity and
  quality of online citizen science participation?}
\newblock \bibinfo{journal}{{\em PloS one\/}} \bibinfo{volume}{9},
  \bibinfo{number}{4} (\bibinfo{year}{2014}), \bibinfo{pages}{e90375}.
\newblock


\bibitem[\protect\citeauthoryear{Raddick, Bracey, Gay, Lintott, Murray,
  Schawinski, Szalay, and Vandenberg}{Raddick et~al\mbox{.}}{2010}]%
        {raddick2009galaxy}
\bibfield{author}{\bibinfo{person}{Jordan Raddick}, \bibinfo{person}{Georgia
  Bracey}, \bibinfo{person}{Pamela~L Gay}, \bibinfo{person}{Chris~J Lintott},
  \bibinfo{person}{Phil Murray}, \bibinfo{person}{Kevin Schawinski},
  \bibinfo{person}{Alexander~S Szalay}, {and} \bibinfo{person}{Jan
  Vandenberg}.} \bibinfo{year}{2010}\natexlab{}.
\newblock \showarticletitle{Galaxy Zoo: Exploring the Motivations of Citizen
  Science Volunteers}.
\newblock \bibinfo{journal}{{\em Astronomy Education Review\/}}
  \bibinfo{volume}{9}, \bibinfo{number}{1} (\bibinfo{year}{2010}).
\newblock


\bibitem[\protect\citeauthoryear{Reeves, West, and Simperl}{Reeves
  et~al\mbox{.}}{2018}]%
        {soton419303}
\bibfield{author}{\bibinfo{person}{Neal Reeves}, \bibinfo{person}{Peter West},
  {and} \bibinfo{person}{Elena Simperl}.} \bibinfo{year}{2018}\natexlab{}.
\newblock \showarticletitle{"A game without competition is hardly a game'': The
  impact of competitions on player activity in a human computation game}. In
  \bibinfo{booktitle}{{\em Proceedings of the Sixth AAAI Conference on Human
  Computation and Crowdsourcing (HCOMP-18)}}. \bibinfo{publisher}{AAAI}.
\newblock


\bibitem[\protect\citeauthoryear{Ritter, Clark, Etzioni, et~al\mbox{.}}{Ritter
  et~al\mbox{.}}{2011}]%
        {ritter2011named}
\bibfield{author}{\bibinfo{person}{Alan Ritter}, \bibinfo{person}{Sam Clark},
  \bibinfo{person}{Oren Etzioni}, {et~al\mbox{.}}}
  \bibinfo{year}{2011}\natexlab{}.
\newblock \showarticletitle{Named entity recognition in tweets: an experimental
  study}. In \bibinfo{booktitle}{{\em Proceedings of the Conference on
  Empirical Methods in Natural Language Processing}}. Association for
  Computational Linguistics, \bibinfo{pages}{1524--1534}.
\newblock


\bibitem[\protect\citeauthoryear{Rokicki, Chelaru, Zerr, and
  Siersdorfer}{Rokicki et~al\mbox{.}}{2014}]%
        {rokicki2014competitive}
\bibfield{author}{\bibinfo{person}{Markus Rokicki}, \bibinfo{person}{Sergiu
  Chelaru}, \bibinfo{person}{Sergej Zerr}, {and} \bibinfo{person}{Stefan
  Siersdorfer}.} \bibinfo{year}{2014}\natexlab{}.
\newblock \showarticletitle{Competitive game designs for improving the cost
  effectiveness of crowdsourcing}. In \bibinfo{booktitle}{{\em Proceedings of
  the 23rd ACM International Conference on Conference on Information and
  Knowledge Management}}. ACM, \bibinfo{pages}{1469--1478}.
\newblock


\bibitem[\protect\citeauthoryear{Rokicki, Zerr, and Siersdorfer}{Rokicki
  et~al\mbox{.}}{2015}]%
        {rokicki2015groupsourcing}
\bibfield{author}{\bibinfo{person}{Markus Rokicki}, \bibinfo{person}{Sergej
  Zerr}, {and} \bibinfo{person}{Stefan Siersdorfer}.}
  \bibinfo{year}{2015}\natexlab{}.
\newblock \showarticletitle{Groupsourcing: Team competition designs for
  crowdsourcing}. In \bibinfo{booktitle}{{\em Proceedings of the 24th
  International Conference on World Wide Web}}. \bibinfo{pages}{906--915}.
\newblock


\bibitem[\protect\citeauthoryear{Rughinis}{Rughinis}{2013}]%
        {rughinis2013gamification}
\bibfield{author}{\bibinfo{person}{Razvan Rughinis}.}
  \bibinfo{year}{2013}\natexlab{}.
\newblock \showarticletitle{Gamification for productive interaction: Reading
  and working with the gamification debate in education}. In
  \bibinfo{booktitle}{{\em Information Systems and Technologies (CISTI), 2013
  8th Iberian Conference on}}. IEEE, \bibinfo{pages}{1--5}.
\newblock


\bibitem[\protect\citeauthoryear{Seaborn and Fels}{Seaborn and Fels}{2015}]%
        {seaborn2015gamification}
\bibfield{author}{\bibinfo{person}{Katie Seaborn} {and}
  \bibinfo{person}{Deborah~I Fels}.} \bibinfo{year}{2015}\natexlab{}.
\newblock \showarticletitle{Gamification in theory and action: A survey}.
\newblock \bibinfo{journal}{{\em International Journal of Human-Computer
  Studies\/}}  \bibinfo{volume}{74} (\bibinfo{year}{2015}),
  \bibinfo{pages}{14--31}.
\newblock


\bibitem[\protect\citeauthoryear{Settles}{Settles}{2012}]%
        {settles2012active}
\bibfield{author}{\bibinfo{person}{Burr Settles}.}
  \bibinfo{year}{2012}\natexlab{}.
\newblock \showarticletitle{Active learning}.
\newblock \bibinfo{journal}{{\em Synthesis Lectures on Artificial Intelligence
  and Machine Learning\/}} \bibinfo{volume}{6}, \bibinfo{number}{1}
  (\bibinfo{year}{2012}), \bibinfo{pages}{1--114}.
\newblock


\bibitem[\protect\citeauthoryear{Sheng, Provost, and Ipeirotis}{Sheng
  et~al\mbox{.}}{2008}]%
        {sheng2008get}
\bibfield{author}{\bibinfo{person}{Victor~S Sheng}, \bibinfo{person}{Foster
  Provost}, {and} \bibinfo{person}{Panagiotis~G Ipeirotis}.}
  \bibinfo{year}{2008}\natexlab{}.
\newblock \showarticletitle{Get another label? improving data quality and data
  mining using multiple, noisy labelers}. In \bibinfo{booktitle}{{\em
  Proceedings of the 14th ACM SIGKDD International Conference on Knowledge
  Discovery and Data Mining}}. ACM, \bibinfo{pages}{614--622}.
\newblock


\bibitem[\protect\citeauthoryear{Tang, Cebrian, Giacobe, Kim, Kim, and
  Wickert}{Tang et~al\mbox{.}}{2011}]%
        {tang2011reflecting}
\bibfield{author}{\bibinfo{person}{John~C Tang}, \bibinfo{person}{Manuel
  Cebrian}, \bibinfo{person}{Nicklaus~A Giacobe}, \bibinfo{person}{Hyun-Woo
  Kim}, \bibinfo{person}{Taemie Kim}, {and} \bibinfo{person}{Douglas~Beaker
  Wickert}.} \bibinfo{year}{2011}\natexlab{}.
\newblock \showarticletitle{Reflecting on the DARPA red balloon challenge}.
\newblock \bibinfo{journal}{{\it Commun. ACM}} \bibinfo{volume}{54},
  \bibinfo{number}{4} (\bibinfo{year}{2011}), \bibinfo{pages}{78--85}.
\newblock


\bibitem[\protect\citeauthoryear{Tinati, Luczak-Roesch, Simperl, and
  Hall}{Tinati et~al\mbox{.}}{2017}]%
        {tinati2017investigation}
\bibfield{author}{\bibinfo{person}{Ramine Tinati}, \bibinfo{person}{Markus
  Luczak-Roesch}, \bibinfo{person}{Elena Simperl}, {and} \bibinfo{person}{Wendy
  Hall}.} \bibinfo{year}{2017}\natexlab{}.
\newblock \showarticletitle{An investigation of player motivations in Eyewire,
  a gamified citizen science project}.
\newblock \bibinfo{journal}{{\em Computers in Human Behavior\/}}
  \bibinfo{volume}{73} (\bibinfo{year}{2017}), \bibinfo{pages}{527--540}.
\newblock


\bibitem[\protect\citeauthoryear{Vijayanarasimhan and Grauman}{Vijayanarasimhan
  and Grauman}{2014}]%
        {vijayanarasimhan2014large}
\bibfield{author}{\bibinfo{person}{Sudheendra Vijayanarasimhan} {and}
  \bibinfo{person}{Kristen Grauman}.} \bibinfo{year}{2014}\natexlab{}.
\newblock \showarticletitle{Large-scale live active learning: Training object
  detectors with crawled data and crowds}.
\newblock \bibinfo{journal}{{\em International Journal of Computer Vision\/}}
  \bibinfo{volume}{108}, \bibinfo{number}{1-2} (\bibinfo{year}{2014}),
  \bibinfo{pages}{97--114}.
\newblock


\bibitem[\protect\citeauthoryear{von Ahn and Dabbish}{von Ahn and
  Dabbish}{2004}]%
        {vonAhn:2004:LIC:985692.985733}
\bibfield{author}{\bibinfo{person}{Luis von Ahn} {and} \bibinfo{person}{Laura
  Dabbish}.} \bibinfo{year}{2004}\natexlab{}.
\newblock \showarticletitle{Labeling Images with a Computer Game}. In
  \bibinfo{booktitle}{{\em Proceedings of the SIGCHI Conference on Human
  Factors in Computing Systems}} {\em (\bibinfo{series}{CHI '04})}.
  \bibinfo{publisher}{ACM}, \bibinfo{address}{New York, NY, USA},
  \bibinfo{pages}{319--326}.
\newblock
\showISBNx{1-58113-702-8}


\bibitem[\protect\citeauthoryear{Von~Ahn and Dabbish}{Von~Ahn and
  Dabbish}{2008}]%
        {von2008designing}
\bibfield{author}{\bibinfo{person}{Luis Von~Ahn} {and} \bibinfo{person}{Laura
  Dabbish}.} \bibinfo{year}{2008}\natexlab{}.
\newblock \showarticletitle{Designing games with a purpose}.
\newblock \bibinfo{journal}{{\it Commun. ACM}} \bibinfo{volume}{51},
  \bibinfo{number}{8} (\bibinfo{year}{2008}), \bibinfo{pages}{58--67}.
\newblock


\bibitem[\protect\citeauthoryear{Xie, Yang, Vasilakos, and He}{Xie
  et~al\mbox{.}}{2014}]%
        {xie2014fair}
\bibfield{author}{\bibinfo{person}{Xianzhong Xie}, \bibinfo{person}{Helin
  Yang}, \bibinfo{person}{Athanasios~V Vasilakos}, {and} \bibinfo{person}{Lu
  He}.} \bibinfo{year}{2014}\natexlab{}.
\newblock \showarticletitle{Fair power control using game theory with pricing
  scheme in cognitive radio networks}.
\newblock \bibinfo{journal}{{\em Journal of Communications and Networks\/}}
  \bibinfo{volume}{16}, \bibinfo{number}{2} (\bibinfo{year}{2014}),
  \bibinfo{pages}{183--192}.
\newblock


\bibitem[\protect\citeauthoryear{Yan, Rosales, Fung, and Dy}{Yan
  et~al\mbox{.}}{2011}]%
        {yan2011active}
\bibfield{author}{\bibinfo{person}{Yan Yan}, \bibinfo{person}{Romer Rosales},
  \bibinfo{person}{Glenn Fung}, {and} \bibinfo{person}{Jennifer~G Dy}.}
  \bibinfo{year}{2011}\natexlab{}.
\newblock \showarticletitle{Active learning from crowds.}. In
  \bibinfo{booktitle}{{\em ICML}}, Vol.~\bibinfo{volume}{11}.
  \bibinfo{pages}{1161--1168}.
\newblock


\bibitem[\protect\citeauthoryear{Yang, Fang, and Xue}{Yang
  et~al\mbox{.}}{2012}]%
        {yang2012game}
\bibfield{author}{\bibinfo{person}{Dejun Yang}, \bibinfo{person}{Xi Fang},
  {and} \bibinfo{person}{Guoliang Xue}.} \bibinfo{year}{2012}\natexlab{}.
\newblock \showarticletitle{Game theory in cooperative communications}.
\newblock \bibinfo{journal}{{\em IEEE Wireless Communications\/}}
  \bibinfo{volume}{19}, \bibinfo{number}{2} (\bibinfo{year}{2012}).
\newblock


\bibitem[\protect\citeauthoryear{Zichermann}{Zichermann}{2011}]%
        {zichermann2011gamification}
\bibfield{author}{\bibinfo{person}{Gabe Zichermann}.}
  \bibinfo{year}{2011}\natexlab{}.
\newblock \bibinfo{title}{Gamification has issues, but they aren't the ones
  everyone focuses on}.
\newblock   (\bibinfo{year}{2011}).
\newblock


\end{thebibliography}

\end{document}